\documentstyle[twocolumn,aps]{revtex}

\begin{document}
\def\vec#1{{\bf#1}}
\def\hatn#1{\hat{\bf#1}}

%\titlepage
\thispagestyle{empty}

\bigskip

\title{Finite-Wavevector Electromagnetic Response
 of Fractional Quantized Hall States}
\author{Steven H. Simon and Bertrand I. Halperin}

\bigskip

\address{Physics Department, Harvard University, Cambridge MA 02138  \\
\vspace{.3in}\begin{minipage}[t]{7in} \rm
A fractional quantized Hall state with filling fraction $\nu =
p/(2mp+1)$ can be modeled as an integer quantized Hall state of
transformed fermions, interacting with a Chern-Simons field. The
electromagnetic response function for these states at arbitrary
frequency and wavevector can be calculated using a semiclassical
approximation or the Random Phase Approximation (RPA).  However,
such calculations do not properly take into account the large
effective mass renormalization which is present in the Chern-Simons
theory.  We show how the mass renormalization can be incorporated
in a calculation of the response function within a Landau Fermi
liquid theory approach such that Kohn's theorem and the $f$-sum
rules are  properly satisfied.  We present results of such
calculations.\end{minipage}  }

\maketitle

\bigskip

\section{Introduction}
\label{sec:intro}

Although the ground state for fractional quantized Hall systems
is reasonably well understood at the Laughlin filling fractions
$\nu =1/(2m+1)$ where $m$ is an integer, we have only a
qualitative understanding of the elementary excitations of the
system\cite{Book}.  Furthermore, theories of experimentally
observed fractional quantized Hall states at other filling
fractions remain controversial.  In order to understand the more
general series of states at filling fractions $\nu=p/(2mp+1)$
where $m$ and $p$ are integers, Jain has constructed trial
wavefunctions based on a picture of ``composite fermions,'' which
may be described loosely as electrons bound to an even number of
magnetic flux quanta\cite{Jain}.  The fractional quantized states
correspond to integer quantized Hall states for the composite
fermions in Jain's description.  Lopez and Fradkin\cite{Lopez1}
showed how one can formally transform the electron system at
$\nu=p/(2mp+1)$ into a system of fermions interacting with a
Chern-Simons gauge field, such that in the mean-field
approximation the ground state is indeed a system of $p$ filled
Landau levels for the transformed fermions, in accord with Jain's
analysis.  Moreover, Lopez and Fradkin proposed that, going
beyond mean-field theory, one could employ the Random Phase
Approximation (RPA) or time-dependent Hartree approximation to
calculate the linear response functions to an external
electromagnetic field at wavevector $\vec q$ and frequency
$\omega$.   This calculation was carried out\cite{Lopez2} to
obtain an optical excitation spectrum for these quantized Hall
states in the limit of $q \rightarrow 0$.  The
fermion-Chern-Simons picture of Lopez and Fradkin was further
developed by Halperin, Lee, and Read (henceforth referred to as
HLR\cite{Tome}) who used it to study even-denominator filling
fractions such as $\nu=1/2$, where no quantized Hall effect is
observed.  Their analysis also had implications for the
excitation spectra of Jain's quantized Hall states, especially in
the limit $p \rightarrow \infty$, where the value of
 $\nu$
approaches an even fraction.

An important correction to both the mean-field theory and the
RPA, noted by HLR, is that fluctuations in the Chern-Simons gauge
field lead to a large correction to the effective mass $m^*$ that
describes low energy excitations.  The RPA, in its standard form,
as used by Lopez and Fradkin, assumes an effective mass which is
equal to the bare electron band mass $m_b$.  If one arbitrarily
changes the value of the mass in the RPA in order to get
reasonable energies for the lowest branch of the excitation
spectrum for the fractional quantized Hall states, then one
violates both the $f$-sum rule and Kohn's theorem, which says
that in the limit $q \rightarrow 0$ a mode at the bare cyclotron
frequency $\omega_c = |eB|/(m_b c)$ has all the weight of the
$f$-sum rule.  In the present paper  we propose a modification of
the RPA which we believe gives a good representation of the low
energy branches of the spectrum, while at the same time
preserving the $f$-sum rule when $m^* \ne m_b$.  Our modified RPA
may be obtained as a natural extension of Landau-Silin Fermi
liquid theory if one includes in addition to the direct Coulomb
potential and the self consistent Chern-Simons field a nonzero
value of the Fermi liquid coefficient $A_1$, chosen to satisfy
the constraint imposed by Galilean invariance.  We present
numerical results of this approximation for a wide range of $q$
values for three representative quantized Hall states ($\nu
=\frac{1}{3}, \, \frac{3}{7}, \, \frac{10}{21}$), comparing the
results with those of the unrenormalized RPA and a semiclassical
approximation suggested by HLR.

Although the electromagnetic response of the quantized Hall state
is trivial in the absence of impurities at zero wavevector
($q=0$) because of Kohn's theorem, the finite wavevector
excitation spectrum can display a very rich structure.
Theoretical calculations of such excitation spectra have been
accurately performed in a controlled perturbation theory  only
for integer quantized Hall states\cite{Kallin}.  The spectra for
fractional quantized Hall states have been much harder to
calculate.   To this end, Girvin, Macdonald, and
Platzman\cite{Girvin1} used a single-mode approximation in
analogy with the Feynman theory of superfluid helium to determine
the dispersion relation of the lowest energy branch of the
excitation spectrum of the Laughlin states $\nu=1/(2m+1)$.  In
this approximation, it was shown that there is a gap at zero
wavevector, and a minimum in the dispersion curve at finite
wavevector.   This minimum was called the ``magnetoroton'' in
analogy with  superfluid helium.

Excitation spectra have also been calculated exactly by numerical
diagonalization of small spherical systems restricted to the
first Landau level\cite{SongHe,Morf}.  The results obtained in
the present paper are in at least reasonable qualitative
agreement with available exact calculations of the density
response function for these finite systems\cite{SongHe}. (A
detailed comparison will be given elsewhere\cite{Me}).  A
particularly interesting feature of our results is that for large
$p$, the lowest branch in the excitation spectrum acquires a
series of deep minima, similar to the magnetoroton minimum at
$\nu=1/3$, at wavevectors given approximately by $q_n \approx
k_{\mbox{\tiny F}}(n+\frac{1}{4}) \pi /(2|p|)$ where $n$ is an
integer, and $ k_{\mbox{\tiny F}}$ is related to the electron
density $n_e$ by $ k_{\mbox{\tiny F}} = (4 \pi n_e)^{\frac{1}{2}}
$.

It should also be possible to experimentally observe the finite
wavevector excitation spectrum via resonant inelastic light
scattering with an angle of incidence far from the normal.  In
fact, such measurements have been performed recently on integer
quantized Hall states\cite{Raman}. It may also be possible to use
a grating near the surface of the quantized Hall system to
measure the electromagnetic response at the wavevector of the
grating.  Comparison of the results of these experiments to our
present calculations should provide an excellent test of our
current understanding of the fractional quantized Hall effect.

The outline of this paper is as follows. In Sec. \ref{sec:Model}
we review the model used by HLR\cite{Tome}.  We describe the RPA
and the semiclassical approximations.  In these models, an
unperturbed response function is calculated and the interactions
as well as the gauge fluctuations are accounted for in
perturbation theory (essentially by summing bubble diagrams).  In
the RPA, the unperturbed response function is simply the response
function for the mean-field system, whereas for the semiclassical
approximation the unperturbed response function is determined
from a semiclassical approximation of the quasiparticle
conductivity. These unperturbed response functions  are
calculated in Sec. \ref{sec:mean}.  The results of these
calculations are put in more usable form in Appendix
\ref{app:RPAsum}  and Appendix \ref{app:semisum}.  In Sec.
\ref{sec:mstar} we discuss the issue of the renormalization of
the quasiparticle mass.  A Fermi liquid theory approach is used
to account for the renormalized mass in the calculation of the
electromagnetic response.   Using this approach, we construct
what we call the ``modified semiclassical'' approximation and the
``modified RPA.''    A more general Fermi liquid theory
calculation that can be generalized to include the effects of
additional nonzero Fermi liquid coefficients is performed in
Appendix \ref{app:FLT} and agrees with the results of Sec.
\ref{sec:mstar}.  In Sec.  \ref{sec:res} we display
electromagnetic response spectra for the quantized Hall states
$\nu =\frac{1}{3}, \, \frac{3}{7}, \, \frac{10}{21}$ calculated
in the semiclassical approximation, the RPA, and the modified
RPA.  Finally, we summarize our findings in Sec.  \ref{sec:summ}.

\section{Model}
\label{sec:Model}

We consider a two-dimensional system of spinless electrons of band
mass $m_b$ and charge $-e$, with interactions given by a potential
$v(r)$, in a uniform magnetic field $B$ perpendicular to the plane
of the system (in the $\hatn z$ direction).  We will generally
take the interaction potential to be the physically interesting
Coulomb interaction given by
\begin{equation}
\label{eq:coul}
v(r) = \frac{e^2}{\epsilon |r|}
\end{equation}
where $\epsilon$ is the background dielectric constant.

Following HLR\cite{Tome}, we  make a singular gauge transformation
to write the Hamiltonian for this system in terms of the composite
fermion quasiparticle creation operator $\psi^+(\vec r)$ that
creates an electron at point $\vec r$ bound to $\tilde \phi$
quanta of Chern-Simons flux.  In terms of these quasiparticle
operators, the Hamiltonian for this system can be written exactly
as
\begin{equation}
    \label{eq:firstHam}
    H=K+V
\end{equation}
where $K$  is the kinetic energy given by
\begin{equation}
K=\frac{\hbar^2}{2m_b}\int d^2 r \hspace{4pt} \psi^+(\vec r)
\left[ -i\vec\nabla +
\{  \frac{e}{c} \vec
A(\vec r)-\vec a(\vec r) \} \right]^2 \psi(\vec r)_.
\end{equation}
 and $V$ is the potential energy
\begin{equation}
V=\frac{1}{2} \int d^2 r\int d^2 r' \hspace{4pt} v(\vec r - \vec
r\,') :\rho(\vec r) \rho(\vec r\,'):
\end{equation}

Here, the colons represent normal ordering of the creation and
annihilation operators, $\vec A(\vec r)$ is the vector potential
due to the
magnetic field $B$ such that $\vec \nabla \times \vec A = B$, and
$\vec a(\vec r)$ is the vector potential associated with the
Chern-Simons flux which can be written as
\begin{eqnarray}
\vec a(\vec r) &=& \tilde \phi \int d^2 r' \vec g(\vec r
 - \vec r\,')
\rho(\vec r\,')
\\
\vec g(\vec r) &=& (\hatn z \times \vec r)/r^2
\end{eqnarray}
where $\hatn z$ is the unit vector perpendicular to the plane of
the system and $\tilde \phi = 2m$ is the even number of flux
quanta bound to each electron.  The point $\vec r = \vec r\,'$
should be excluded from the  Green's function $\vec g(\vec r -
\vec r\,')$.

If we use a mean-field description and average the effect of the
fluctuating gauge field $\vec a$, the Hamiltonian
(\ref{eq:firstHam}) simply represents quasiparticle fermions in a
magnetic field
\begin{equation}
\label{eq:DeltaB}
\Delta B = B-B_{(1/2m)}= B - \frac{4m\pi  \hbar c
n_e}{e}
\end{equation}
where $n_e$ is the electron number density.  With the idea of
perturbing around this mean-field description to account for gauge
fluctuations and interactions, we write the mean-field reference
Hamiltonian as
\begin{equation}
\label{eq:H0}
H_0 =
\frac{\hbar^2}{2m_b} \int d^2 r
\hspace{4pt} \psi^+(\vec r) \left[-i\vec \nabla +
\frac{e}{c} \Delta
\vec A(\vec r)\right]^2 \psi(\vec r)
\end{equation}
 where  the mean-field vector potential
$\Delta \vec A(\vec r)$  satisfies $
\vec \nabla \times (\Delta \vec A) =\Delta B$.

Since the mean-field Hamiltonian describes fermions in a magnetic
field $\Delta B$, the energy levels are simply the usual Landau
levels, but they are now the energy levels of the quasiparticle
wavefunctions.  If  we add a small amount of disorder to the
system, we expect to see the integer quantized Hall effect for the
quasiparticles where the steps are centered around integer
quasiparticle filling fractions.  Since the integer quantized Hall
effect for electrons occurs when the filling fraction $ \nu = n_e
2 \pi \hbar c / (e B)$ is an integer, we now have the integer
quantized Hall effect for {\it quasiparticles} when $p = n_e 2 \pi
\hbar c  /(e \Delta B)$ is an integer.  Substituting in the
definition (\ref{eq:DeltaB}) of $\Delta B$ now yields stable
states at
\begin{eqnarray}
    B&=&B_{(1/2m)}+ \frac{2\pi\hbar c n_e}{e p}
\\
     &=& \frac{2\pi\hbar c n_e}{e} \left[ 2m + \frac{1}{p} \right]
\end{eqnarray}
which corresponds to the Jain states $\nu = p/(2mp+1)$.

We now define the electromagnetic linear response function $K_{\mu
\nu}(\vec q, \omega)$ where $\mu$ and $\nu$ take on the values
$(0,x,y)$ by the relation \begin{equation} j_{\mu}(\vec q, \omega)
= \frac{e}{c} K_{\mu \nu}(\vec q, \omega)
A_{\nu}^{\mbox{\scriptsize{ext}}}(\vec q, \omega) \end{equation}
where $A_{\nu}^{\mbox{\scriptsize{ext}}}$ is an external
perturbing scalar ($\nu = 0$) or vector  ($\nu = x,y$) potential
with frequency $\omega$ and wavevector $\vec q$, and $j_\mu$ is
the induced change in the particle density ($\mu=0$) or current
($\mu = x,y$).  Following HLR\cite{Tome} again, we choose $\vec q
\, \| \hatn x$, and we work in the Coulomb gauge so that the
longitudinal part of $\vec A$ is zero (ie, so that $A_x=0$).  With
these choices, the longitudinal part of $\vec j$ is simply
$(\omega/q) j_0$.  Thus we can consider $K_{\mu \nu}$ as a $2
\times 2$ matrix in which the indices take on the values $0$ and
$1$ where the index $1$ indicates the transverse or $\hatn
y$-direction.

The response function can be calculated within the RPA or
time-dependent Hartree approximation in analogy  with recent work
on anyon superconductivity\cite{Fetter,YHChen,Dai}.  In this work,
the RPA equations are derived through a Hamiltonian formalism.
Alternatively one can derive the same relations from a more
field-theoretic Lagrangian approach\cite{Lopez1,Tome}.   It should
be noted that the RPA formalism of the above mentioned works on
anyon superconductivity\cite{Fetter,YHChen,Dai} differs slightly
from the formalism of HLR\cite{Tome} that we have chosen to
follow.  In particular the HLR formalism is simplified because the
diamagnetic term is included in the bare response.  None-the-less,
both approaches give the same final results.  It should also be
noted that the above mentioned formalisms of Refs.
\cite{Fetter,YHChen,Dai}, as well as that of Lopez and
Fradkin\cite{Lopez1,Lopez2}, involve the calculation of a $3
\times 3$ response matrix, whereas the HLR approach uses a
convenient gauge to reduce the problem to the calculation of a $2
\times 2$ matrix.

The RPA equation for the electromagnetic
 response function is given by
\begin{equation}
\label{eq:RPA}
    K = K^0[1+UK^0]^{-1}
\end{equation}
where $K^0_{\mu \nu}$ is the
response function for the non-interacting system
 of quasiparticles governed by the
 Hamiltonian $H_0$, and the interaction
matrix $U$ is given by
\begin{equation}
    \label{eq:Udef}
    U=V+C^{-1}
\end{equation}
where
\begin{equation}
C=\frac{e^2}{2\pi \hbar \tilde \phi}
\left[ \begin{array}{cc} 0 & iq \\
                        -iq & 0 \end{array} \right]
\end{equation}
is the Chern-Simons interaction and
\begin{equation}
V=\left[ \begin{array}{cc} v(q) & 0 \\
                         0 & 0 \end{array} \right]
\end{equation}
represents the interaction of the quasiparticles through the
potential $v(q)$ which is just the Fourier transform of the
potential $v(r)$.  For the physically relevant case of the Coulomb
potential (Eq. (\ref{eq:coul})) we have \begin{equation} v(q) =
\frac{2\pi e^2}{\epsilon q}.  \end{equation} Note that the
potential $V$  couples the density of particles to the scalar
potential, whereas the Chern-Simons interaction $C$ -- like a
magnetic field -- couples to the current also.

It is sometimes more useful to think in terms of the conductivity
rather than the electromagnetic response\cite{Tome}.  The
conductivity $\sigma$ is defined as the response to the total
electromagnetic field  $A_\mu$ whereas the electromagnetic
response $K$ is the response to the external electromagnetic field
$A_{\mu}^{\mbox{\scriptsize{ext}}}$.  The magnetic field generated
by the quantum Hall system is small, so there is essentially no
difference between $\vec A$ and $\vec
A^{\mbox{\scriptsize{ext}}}$.  On the other hand, the scalar
potentials in Fourier space  $eA_0(q)$ and
$eA_0^{\mbox{\scriptsize{ext}}} (q)$ differ by the Coulomb
potential $v(q)j_0(q)$ generated by the density fluctuations.
Thus, we will define a $2 \times 2$ matrix $\Pi(q,\omega)$, which
is more closely related to the conductivity, to be the
electromagnetic response without this Coulomb contribution:
\begin{equation}
    K^{-1} =  \Pi^{-1} + V \label{eq:Pidef} .
\end{equation}
Similarly, it is convenient to define a
 $2 \times 2$ matrix $\tilde K(q,\omega)$ to be
the electromagnetic response without the Coulomb
 contribution or the
Chern-Simons contribution:
\begin{equation}
\\  \label{eq:Ktildedef}
       K^{-1}      = \tilde K^{-1} + U  .
\end{equation}
In other words, $\Pi$ is the contribution from all Feynman
diagrams for $K$ that are irreducible with respect to $V$ and
$\tilde K$ is the sum of all diagrams for $K$ that are irreducible
with respect to $U$.  Note that the RPA equation (\ref{eq:RPA}) is
obtained by simply approximating $\tilde K$ as the mean-field
noninteracting quasiparticle  response function $K^0$.  The
perturbation $U$ -- which includes the Coulomb and Chern-Simons
interactions -- is then incorporated in Eq.  (\ref{eq:Ktildedef})
to give the full response function.  Similarly, in our
semiclassical approximation we will directly try to approximate
the unperturbed response function  $\tilde K$ for the
quasiparticles.

Maintaining our convention that
 $\vec q\,  \| \hatn x$ we can now follow
HLR\cite{Tome} to define the
conductivity tensor $\sigma_{ij}(q,\omega)$ as
\begin{eqnarray}
\sigma_{xx}^{-1} (q,\omega) &=& \frac{iq^2}{\omega} \left[
\Pi^{-1}_{00}(q,\omega)
-\Pi^{-1}_{00}(q,0) \right] \\
\sigma_{yy}(q,\omega) &=&\frac{-i}{\omega}\left[ \Pi_{11}(q,\omega)
-\Pi_{11}(q,0)
\right] \\
\sigma_{xy}(q,\omega) &=&
-\sigma_{yx}(q,\omega) = \frac{i}{q} \Pi_{01}(q,\omega).
\end{eqnarray}
Similarly, we can define the
``quasiparticle conductivity tensor'' $\tilde
\sigma_{ij} (q,\omega)$ as
\begin{eqnarray}
\tilde \sigma_{xx}^{-1} (q,\omega) &=& \frac{iq^2}{\omega} \left[
\tilde K^{-1}_{00}(q,\omega)
-\tilde K^{-1}_{00}(q,0) \right] \\
\tilde \sigma_{yy}(q,\omega)
 &=&\frac{-i}{\omega}\left[ \tilde K_{11}
(q,\omega)
-\tilde K_{11}(q,0)
\right] \\
\tilde \sigma_{xy}(q,\omega) &=&
-\tilde \sigma_{yx}(q,\omega) =
\frac{i}{q} \tilde K_{01}(q,\omega).
\end{eqnarray}
These definitions have been chosen so that the conductivities are
finite in the $\omega \rightarrow 0$ limit for any fixed value of
$q$.  Although these are not necessarily the only such definitions
that are possible, we will not be overly concerned with the low
frequency limit in this paper.  In fact the contributions from the
zero frequency parts of these relationships are suspected to be
negligibly small in all cases that we will consider.  Thus, we can
approximate these relationships by dropping the additive zero
frequency pieces to write the results in a convenient matrix form
as
\begin{eqnarray}
    \sigma &=& T\Pi T \label{eq:TPi}
    \\
    \tilde \sigma &=& T \tilde K T  \label{eq:TKtilde}
\end{eqnarray}
where $T$ is the conversion matrix
\begin{equation}
T=
\left[ \begin{array}{cc} \frac{i\sqrt{i\omega}}{q} & 0 \\
                         0 & \frac{1}{\sqrt{i\omega}}
\end{array} \right]
\end{equation}
Now from these relations
% (Eqs. (\ref{eq:TPi}) and (\ref{eq:TKtilde}) )
and the
definitions of $\Pi$, $\tilde K$ and $U$
%(Eqs.  (\ref{eq:Pidef}),
%(\ref{eq:Ktildedef}), and (\ref{eq:Udef}) )
we can derive
\begin{eqnarray}
    \label{eq:SemRPA1}
    K^{-1} &=& T^{-1} \rho T^{-1} + V,
\\
    \label{eq:SemRPA2}
    \rho&=&\tilde \rho + \rho_{cs}, \\
\rho_{cs}& \equiv &
    T^{-1} C^{-1} T^{-1} = \frac{2\pi \hbar \tilde \phi} { e^2   }
                \left[ \begin{array}{cc} 0 & 1  \\
                          -1 & 0 \end{array} \right]_,
\end{eqnarray}
where $\rho_{cs}$ is the contribution from the Chern-Simons
interaction,  and $\rho$ and $\tilde\rho$ are the
associated resistivity
 matrices defined as
\begin{eqnarray}  \label{eq:rfroms}
   \rho &=& \sigma^{-1}
\\   \label{eq:rtfromst}
    \tilde \rho &=& \tilde \sigma^{-1}.
\end{eqnarray}
  Note
that if we approximate $\tilde \sigma$ by the mean-field
noninteracting
result $TK^0T$, then Eqs. (\ref{eq:SemRPA1}) and (\ref{eq:SemRPA2})
become equivalent to the RPA prescription  (Eq. (\ref{eq:RPA})).

\section{Unperturbed Response}
\label{sec:mean}

In order to calculate the response function in the RPA
approximation we must first find the response $K^0$ of the
unperturbed mean-field Hamiltonian $H_0$.  In the RPA, the
mean-field response $K^0$ is used as an approximation for the
$U$-irreducible diagrams $\tilde K$.  Similarly, for our
semiclassical approximation, the semiclassical quasiparticle
conductivity $\tilde \sigma$ is used as  an approximation for
$T\tilde KT$.  For a noninteracting two-dimensional system of
spinless fermions of mass $m_b$ and charge $-e$  at density $n_e$
in a perpendicular magnetic field $\Delta B$, the magnetic length
$l_{\Delta} $ is defined by
\begin{equation}
    l_{\Delta} = \sqrt{\frac{c\hbar }{e(\Delta B)}}
\end{equation}
and the cyclotron frequency $\Delta \omega_c$ is given by
\begin{equation}
    \label{eq:deltaomegac}
\Delta \omega_c = \frac{e(\Delta B)}{m_b c.}
\end{equation}
The number of ``effective'' quasiparticle Landau levels
filled is given by
\begin{equation}
    p = \frac{2\pi n_e \hbar c}{e(\Delta B)  .}
\end{equation}
It is also convenient to define the dimensionless
reduced wavevector
\begin{equation}
    X=qR_{\Delta} = \frac{2qp}{k_{\mbox{\tiny{F}}}}
\end{equation}
where
\begin{equation}
 R_{\Delta} = \frac{\hbar k_{\mbox{\tiny{F}}} c }{e (\Delta B)}
\end{equation}
is the effective semiclassical cyclotron radius and
$k_{\mbox{\tiny{F}}}=(4\pi n_e)^{\frac{1}{2}}$ is the Fermi
momentum.  We will find that $X$ is a very natural parameter in
terms of which to express our results.  In particular, we will
find that our results approximately scale in terms of $X$ in the
semiclassical (large $p$) limit.   For simplicity of notation, we
will assume from now on that $\Delta B \ge 0$, and hence $p \ge
0$.

\subsection{$K^0$ for the RPA}
\label{sub:K0}

For the simple case where $p$ is an integer, the unperturbed
electromagnetic response function $K^0$ for has been derived  in
connection with the theory of anyon superconductivity by Fetter
et al\cite{Fetter} for the $p=2$ case, and then for general $p$
by Chen et al \cite{YHChen} (and later by Dai et al\cite{Dai}).
The calculation is performed by realizing that the response
function can be related to the ground state expectation value of
the time ordered product of current operators.   This quantity is
then calculated by inserting complete sets of states for free
electrons in a magnetic field.  The final result in matrix form is
\begin{equation}
            \label{eq:K0def}
    K^0 =   T^{-1} \tilde s  T^{-1}
\end{equation}
where the conductivity matrix $\tilde s$
is given by  \cite{YHChen,Dai}
\begin{equation}
            \label{eq:sfree}
      \tilde s  = \frac{pe^2}{2\pi\hbar} \left[ \begin{array}{cc}
                        i \left(\frac{\omega}{\Delta \omega_c}
\right) \Sigma_0
\,\,\,\, & -\Sigma_1 \\
            \Sigma_1 \,\,\,\, & i\left( \frac{\Delta \omega_c}
{\omega} \right)
(\Sigma_2 +1) \end{array}
\right]
\end{equation}
and
\begin{eqnarray}
   \nonumber
  \Sigma_j &=&       \frac{e^{-Y}}{p} \sum_{l=0}^{p-1} \sum_{m=p}^{\infty}
\left\{ \frac{l!}{m!}
\frac{(m-l) Y^{m-l-1} [ L_l^{m-l}(Y)
]^{2-j}}{\left(\frac{\omega+i0^{^+}}{\Delta
 \omega_c}
 \right)^2 - (m-l)^2}  \right. \\
  &  & \left. \left[ (m-l-Y)L_l^{m-l}(Y) +  \label{eq:Sigmadef}
2Y \frac{dL_l^{m-l}(Y)}{dY}\right]^j
\right\}
\end{eqnarray}
where $L^m_l$ is a Laguerre polynomial, and
the expansion parameter $Y$ is given by
\begin{equation}
    Y = \frac{1}{2}(q l_{\Delta })^2 = \frac{p}{4}X^2 .
\end{equation}
It should be noted that the $(l,m)^{th}$ term in the sum
(\ref{eq:Sigmadef}) represents a particle in the $l^{th}$ Landau
level making a virtual transition up to the $m^{th}$ level and
back.  In Appendix \ref{app:RPAsum} we perform some of the above
summations explicitly such that each $\Sigma_j$ is written as a
single sum.  This simplification has proven to be quite useful
for both analytic and numerical analyses.

For noninteger values of $p$ one can interpolate to find the
residues of a given pole in $K^0$ at nonzero frequency.  In terms
of the imaginary part of the response function, we can write
\begin{eqnarray} \nonumber
\mbox{ Im}\left[ K^0_p(\omega \ne 0  )\right] &=&
([p]+1-p) \mbox{ Im} \left[K^0_{[p]}(\omega) \right]
\\ &+& (p - [p])
\mbox{ Im} \left[K^0_{[p]
+1}(\omega)\right]
\end{eqnarray}
where $[p]$ is the greatest integer less or equal to than $p$.
To find the weight of the pole in the response function at zero
frequency, one must use an $f$-sum rule (see also
Sec. \ref{sec:mstar}).  We will, however, limit our attention
 to the integer values of $p$.

\subsection{Semiclassical $\tilde \sigma$}

The other approach we will use is to semiclassically approximate
the quasiparticle conductivity $\tilde \sigma$ and hence $\tilde
K$.  The semiclassical regime is the region where the energy
levels are closely spaced with respect to the other energy scales
of the problem.  This regime occurs at low effective fields
(large $p$) and long wavelength (small $q$) and when $\hbar
\omega$ is much less than the Fermi energy. If we are in this
semiclassical regime  we can consider the quasiparticles as
localized wavepackets moving under the influence of the magnetic
field $\Delta B$ and (as described in Appendix \ref{app:FLT}) we
can approximate the quasiparticle conductivity $\tilde \sigma$ as
\cite{Tome,Harrison}
\begin{equation}
    \label{eq:sigmatilde}
    \tilde\sigma_{ij} = \frac{2i n_e e c}{\Delta B}
            \sum_{n=-\infty}^{\infty}
 \frac{ V_i^{(n)*} V_j^{(n)}}{\frac{\omega}{\Delta \omega_c} - n +
\frac{i}{\tau} }
\end{equation}
where  $\tau$ is the quasiparticle scattering time,
and the velocity coefficients $V_i^{(n)}$, whose
meaning is further elucidated in Appendix \ref{app:FLT},
are defined as
\begin{eqnarray}
    V_{x}^{(n)} &=& \frac{n}{X} J_n(X) \\
    V_{y}^{(n)} &=& -i\frac {dJ_n(X)}{dX}
\end{eqnarray}
where $J_n$ is the $n^{th}$ Bessel function.
 Note that unlike the RPA,  the semiclassical
approximation gives no special significance to integer values of
  $p$.

In this semiclassical approximation the quasiparticle scattering
time $\tau$ is left as a free parameter whereas the above quantum
mechanical calculation of $K^0$ is inherently in the
no-scattering ($\tau \rightarrow \infty$) limit since no
mechanism has been included to account for scattering.  Although
we have the freedom to perform our semiclassical calculations for
finite $\tau$, it is actually more useful to think of the
no-scattering limit.  In this limit poles will appear in the
density-density response function $K_{00}$  at exactly the the
frequencies corresponding to the collective modes of the system.
Furthermore, by taking the $\tau \rightarrow \infty$ limit we can
compare our results with the RPA.   Making this simplification we
can sum the series of Bessel functions exactly to yield a closed
form expression for the quasiparticle conductivity $\tilde
\sigma$.  This sum is performed explicitly  in Appendix
\ref{app:semisum}.

 \section{Effective Mass Renormalization}
\label{sec:mstar}

\subsection{General Considerations}
\label{sub:Gen}

Within the theory considered so far, the quasiparticle effective
mass $m^*$ is just the bare band mass $m_b$.  In this theory we
perturb around a reference Hamiltonian $H_0$ that describes
particles of this unrenormalized mass (Eq.  (\ref{eq:H0})).  We
expect, however,  that the effective mass should be renormalized
by interactions.  In order to estimate the importance of this
mass renormalization, we follow HLR\cite{Tome} to make a crude
estimate of the value of the effective mass.  Assuming that the
electron interaction energy is much less than the spacing between
Landau levels, the  Landau level mixing can be neglected and all
energies of interaction must then be proportional to the
electron-electron interaction energy scale $e^2 (4\pi
n_e)^{1/2}/\epsilon$.  Thus, dimensional analysis tells us that
the effective mass should have the form (if it is in fact finite)
\begin{equation}
    \label{eq:mstarscale}
    m^* = \frac{\hbar^2 (4\pi n_e)^{1/2} \epsilon}{e^2 C}
\end{equation}
where $C$ is a dimensionless constant.  HLR \cite{Tome} use
results from the exact diagonalization of small spherical systems
restricted to the lowest Landau level to estimate that $C=0.3$.
Using the experimentally relevant dielectric constant $\epsilon =
12.6$ appropriate for GaAs, a field of $B=10T$ and a filling
fraction of $\nu = \frac{1}{2}$ yields the result
\begin{equation}
    m^*\approx 4m_b .
\end{equation}

Using a self-consistent analysis of a selected set of diagrams
for the self energy of the transformed fermions which describes
the interaction with long wavelength fluctuations in the
Chern-Simons vector potential, HLR\cite{Tome} conclude that for
the case of the Coulomb interaction between the electrons, the
effective mass $m^*$ should actually exhibit a logarithmic
divergence for energies near the Fermi energy, and for $p
\rightarrow \infty$ (ie, for $\nu \rightarrow \frac{1}{2}$).  The
coefficient in front of the logarithm obtained by HLR is
relatively small, however, and the resulting values of the
effective mass, in practice, will not be very different from
those given by  Eq.  (\ref{eq:mstarscale}).

The important thing to note here is that the mass is renormalized
considerably.  Thus, perturbing around an unrenormalized
Hamiltonian is likely to give very poor results.  The first
na\"{\i}ve thing one could do to try to correct this problem is
simply to use this renormalized effective mass in the reference
Hamiltonian $H_0$.  In fact, in Sec. \ref{sec:res} we will see
that this approach can sometimes give reasonable results for the
dispersion relation of the lowest excitation mode.  However, this
approach will give an incorrect value for the
cyclotron-frequency, thereby violating Kohn's theorem and the
$f$-sum rule.   The focus of this section is the construction of
a method of repairing our na\"{\i}ve approach so that these rules
are satisfied.

First we stop to think about the properties we want our result to
have.  To begin with, we recall Kohn's theorem (a result of
Galilean invariance) requires that the $q \rightarrow 0$ behavior
of our system be determined by the band mass $m_b$ rather than
any renormalized mass.   One can imagine all of the electrons in
the system oscillating in unison so that electron-electron
interactions  have no effect.  Similarly the $f$-sum rule simply
says that the behavior of our system in the $\omega \rightarrow
\infty$ limit is also  determined by the band mass $m_b$.  This
is easily imagined since at high frequency one can think of the
electrons oscillating very quickly with very small magnitude so
that these oscillations do not appreciably change the positions
of the electrons or couple to the electron-electron interaction.
Often this rule is stated in terms of the conductivity such that
by using a Kramers-Kr\"onig relation it can be written as an
integral over frequency (an $f$-sum).

In the long wavelength or high frequency limit the free
electron result is written most easily in terms of the resistivity
\begin{equation}   \label{eq:fsum1}
   \rho \sim \frac{m_b}{e^2 n_e}
   \left[ \begin{array}{cc} -i\omega & \omega_c \\
                         -\omega_c &  -i\omega \end{array} \right]
+ {\cal O}(q^2/\omega).
\end{equation}
 If the resistivity of our system indeed has this high frequency,
long wavelength limit, then Kohn's theorem and the $f$-sum rules
are satisfied.  We now want to turn this condition on the
resistivity (or equivalently the response function) into a
condition on the properties of the quasiparticle system.
Consider the  resistivity  $\tilde \rho$ for quasiparticles with
the band mass $m_b$ in the effective field $\Delta B$.  The
analogous free quasiparticle high frequency, long wavelength
limit is
\begin{equation}
    \label{eq:qpsum}
     \tilde   \rho \sim \frac{m_b}{e^2 n_e}
   \left[ \begin{array}{cc} -i\omega & \Delta \omega_c \\
                       -\Delta  \omega_c &  -i\omega \end{array}
 \right] + {\cal O}(q^2/\omega).
\end{equation}
where $\Delta \omega_c = e\Delta B/(m_b c)$ is the cyclotron
frequency associated with the effective magnetic field and the
band mass. Now if we convert this into the associated resistivity
for the original electron system using Eq. (\ref{eq:SemRPA2}) we
find that
\begin{eqnarray}
    \rho & \sim & \frac{m_b}{e^2 n_e}
   \left[ \begin{array}{cc} -i\omega & \Delta \omega_c \\
                       -\Delta  \omega_c &  -i\omega
\end{array} \right]
 +\frac{2\pi \hbar \tilde \phi} { e^2   }
                \left[ \begin{array}{cc} 0 & 1  \\
                          -1 & 0 \end{array} \right] \\
            &\sim &  \frac{m_b}{e^2 n_e}
   \left[ \begin{array}{cc} -i\omega & \omega_c \\
                         -\omega_c &  -i\omega \end{array} \right]
\end{eqnarray}
where we have made use of the fact (Eq.  \ref{eq:DeltaB}) that
$\Delta B = B -  2 \pi  \hbar c \tilde \phi n_e/e$.  We conclude
that if our quasiparticle system satisfies Kohn's theorem and the
$f$-sum rules with respect to the effective magnetic field
$\Delta B$ and the band mass $m_b$, then our original electron
system satisfies the same rules with respect to the full field
$B$.

 We must now arrange for our {\it quasiparticle}  system to
satisfy Kohn's theorem and the $f$-sum rule when we renormalize
the quasiparticle mass (which we have so far taken to be equal to
the bare band mass $m_b$) to some new value $m^*$.   If we
na\"{\i}vely try to calculate $\tilde \rho$  with the new
renormalized mass, by simply replacing the band mass $m_b$ by the
(phenomenological) effective mass $m^*$ everywhere it occurs, we
must clearly end up with the high frequency, long wavelength
behavior
\begin{equation}
    \label{eq:rhonlim}
     \tilde   \rho^{\mbox{\small n}} \sim  \frac{m^*}{e^2 n_e}
   \left[ \begin{array}{cc} -i\omega & \Delta \omega_c^* \\
               -\Delta  \omega_c^* &  -i\omega \end{array} \right]
+ {\cal O}(q^2).
\end{equation}
where the effective mass renormalized cyclotron frequency
 is defined as
\begin{equation}
        \Delta \omega_c^* = \frac{e \Delta B}{m^* c.}
\end{equation}
This limiting behavior (Eq. (\ref{eq:rhonlim})) is clearly
different from the desired limit given in Eq. (\ref{eq:qpsum}).
Note that the off-diagonal terms are independent of the value of
$m^*$, so only the diagonal terms are in violation of the sum
rules.  We must now find a way to  ``fix'' this result so that
the resistance takes the proper form.  In other words we must
find a way to calculate the quasiparticle conductivity that more
properly incorporates the effective mass and corrects for the
fact that the effective mass is in general not equal to the band
mass.

This problem is now almost exactly the same as the well studied
problem of Fermi liquid effects on magnetoplasma modes in
metals\cite{FermiLiquid,Lee,Ando} where one considers the
excitation modes of electrons in a strong magnetic field.
Theoretically, one can use an approach similar to our
semiclassical calculation of the quasiparticle conductivity to
predict the spectrum of such a system\cite{Harrison}.   Once
again the electron mass is renormalized due to interactions, and
a na\"{\i}ve semiclassical approach will either not account for
this mass renormalization or will violate the sum rules.  One
solution that has been used is to account for the mass
renormalization within a formal Landau-Silin Fermi liquid
theory\cite{FermiLiquid}.  Such a Fermi liquid approach should be
valid at long wavelengths and when $\hbar \omega$ and $\hbar
\omega_c$ are much less than the Fermi energy.  Even more
analogous to our problem, Lee and Quinn have used such a theory
to study two-dimensional electron systems\cite{Lee}.  They show
within this approach that (within the semiclassical regime where
the cyclotron energy as well as $\hbar \omega$ are much less than
the Fermi energy) in the $q \rightarrow 0$ limit the frequency of
the $n^{th}$ excitation mode is given by
\begin{equation}
\omega = (1 + A_n)n \omega_c^*  + {\cal O}(q^2)
\end{equation}
 where $A_n$ is the $n^{th}$ Fermi liquid coefficient.  Note that
since the effective mass is controlled by the first Fermi liquid
coefficient (a result of Galilean invariance of Fermi liquid
theory) via\cite{Pines}
\begin{equation}
    \label{eq:massre}
 m^* = (1+A_1) m_b
\end{equation}
the location of the first ($n=1$)  excitation mode -- the
cyclotron frequency -- is unchanged when the mass is
renormalized.  This is exactly the type of result we want.
Unfortunately, we will need to know the full conductivity, not
just the frequency of the excitation modes, so we will be unable
to use the results of Lee and Quinn directly.  None-the-less, we
will be able to use this type of theory to calculate the
conductivity for quasiparticles with renormalized mass.  One
additional advantage of using this type of Landau Fermi liquid
theory is that we do not need to know exactly how or even why
the electron mass is renormalized, since all of the relevant
details of the electron-electron interaction are  included within
the single mass renormalization coefficient $A_1$.

\subsection{Modified Semiclassical Approximation
for $\tilde \rho$}
\label{sub:ModSem}

We begin by discussing the Fermi liquid corrections to the
quasiparticle resistivity tensor $\tilde \rho$ in the
semiclassical approximation.  To do this, we first ignore the
direct Chern-Simons and Coulomb interactions, and consider the
current induced in the Fermi liquid by a specified
electromagnetic vector potential $\vec A(\vec r,t)$.  Eventually
we shall use the result for $\tilde \rho$ in Eq.
(\ref{eq:SemRPA2}), which will be equivalent to replacing the
electromagnetic field by the sum of the self consistent
Chern-Simons and electromagnetic fields.

 In performing this Fermi liquid theory calculation we make
several simplifying assumptions.  To begin with, as explained
above, we consider only the low scattering $\tau \rightarrow
\infty$ limit.  (The effects of impurity scattering are
considered as the more general case in ref. \cite{Lee}).  We also
assume that higher Fermi liquid coefficients ($A_l$ for $l>1$)
are progressively less important and we can set $A_0=0$ since we
can include the effects of the density-density interaction in the
RPA treatment by modifying the interaction $v(r)$ at short
distances.   Thus,  we assume that all of the Fermi liquid
coefficients are zero except for the coefficient $A_1$ that
controls the mass renormalization via Eq. (\ref{eq:massre}).  In
Appendix \ref{app:FLT} it is shown how to include the effects of
other nonzero Fermi-liquid coefficients.  Finally, in order to
use a Fermi liquid theory, we must assume that we are in the
semiclassical regime where the quasiparticles can be treated as
localized wavepackets.  In other words, we should have the
wavevector $q$ much less than the Fermi wavevector
$k_{\mbox{\tiny{F}}}$ while  $\hbar \omega$ and the spacing
between effective Landau levels $\hbar \Delta \omega_c^*$ must
both be much less than the Fermi energy $E_{\mbox{\tiny F}}$.
The last condition, in particular, is well satisfied for large
$p$, but only marginally satisfied for small $p$ such as at $\nu
= \frac{1}{3}$.

With these assumptions we follow the usual Fermi liquid
approach\cite{Pines} and write the energy functional for a two
dimensional system of spinless quasiparticles of effective mass
$m^*$ at density $n_e$, in an electromagnetic field defined by
the vector potential $\vec A$, as
\begin{eqnarray} \nonumber
    E[n(\vec p, \vec r) ] &=& E_0[n(\vec p, \vec r) ] + \\
 \frac{A_1}{2 n_e m^*} &\sum_{\vec p,\vec p\,'} &
(\vec p+e\vec A)\cdot(\vec p\,'+e\vec A)
 n(\vec p, \vec r)n(\vec p\,', \vec r) \label{eq:FLHam}
\end{eqnarray}
where
\begin{equation}
   E_0[n(\vec p, \vec r) ]  =
    \sum_{\vec p} \frac{(\vec p+e\vec A)^2}{2m^*} n(\vec p, \vec r)
\end{equation}
and $n(\vec p,\vec r)$ is the phase-space density at momentum
$\vec p$ and position $\vec r$.  More generally, as discussed in
Appendix \ref{app:FLT} the energy functional will have additional
interaction  terms with other nonzero Fermi liquid coefficients.
Note that in this section we have set $c=1$ for simplicity.
Equation (\ref{eq:FLHam}) assumes implicitly that $n(\vec p,\vec
r)$ is a slowly varying function of $\vec r$.  We also assume in
this section, that we have chosen a gauge where the scalar
potential is zero.

We can now calculate the local current by differentiating $E$
 with respect to
the vector potential
\begin{equation}
  \vec J = \frac {-\delta E[ n(\vec p,\vec r)]}{\delta \vec A} =
\sum_{\vec p
} \frac{-e(\vec p + e\vec
A)}{m^*} (1+ A_1) n(\vec p,\vec r)
\end{equation}
where we have used the fact that
\begin{equation}
    \sum_{\vec p} n(\vec p,\vec r) = n(\vec r)
\end{equation}
is the local particle density which to lowest order we have taken
 to be equal to
the average particle density $n_e$.  On the other hand, by
Galilean invariance, we expect that
\begin{equation}
    \vec J =  \sum_p \frac{-e(\vec p + e\vec A)}{m_b}
n(\vec p,\vec r).
\end{equation}
By equating these two expressions for the current we easily
derive the relation (\ref{eq:massre}) between bare mass $m_b$ and
effective mass $m^*$.

We are now faced with actually trying to self-consistently
compute the time dependence of the phase-space density $n(\vec
p,\vec r)$ when we apply a perturbing electromagnetic field.  In
order to do this, we begin by constructing an effective single
particle Hamiltonian
\begin{equation}
    \label{eq:Heff}
        H_{\mbox{\small{eff}}}(\vec p, \vec r) =
     \frac{\delta E [n(\vec p,\vec r)]}{\delta n(\vec p,\vec r)}
\end{equation}
we then have Hamilton's equations of motion
\begin{eqnarray}
    \frac{d\vec p}{dt} &=&
-\vec \nabla_{\vec r} H_{\mbox{\small{eff}}}  \\
      \frac{d\vec r }{dt} &=&\vec \nabla_{\vec p}
    H_{\mbox{\small{eff}}}.
\end{eqnarray}
These then are used
to construct the Boltzman (Fokker-Plank) equation of motion
for  $n(\vec p,\vec r)$
\begin{eqnarray}
    \frac{\partial  n(\vec p,\vec r)}{\partial t} &=&
-\left[   \vec \nabla_{\vec p}  \cdot \frac{d\vec p}{dt}+
 \vec \nabla_{\vec r} \cdot \frac{d\vec r }
{dt}
 \right] n(\vec p,\vec r)  \\
&=& \left[ \left( \vec \nabla_{\vec r} H_{\mbox{\small{eff}}}
\right)  \cdot \vec \nabla_{\vec p} -
   \left(    \vec   \nabla_{\vec p} H_{\mbox{\small{eff}}}
\right) \cdot \vec \nabla_{\vec r}
 \right] n(\vec p,\vec r).
\end{eqnarray}
The general self-consistent solution to this equation is
nontrivial, and is outlined in Appendix \ref{app:FLT}. However,
we know the solution of this equation for the simple case where
all of the Fermi liquid interaction terms are set to zero (ie,
there is no mass renormalization $m^* = m_b$).  In this case, the
trivial effective Hamiltonian is just
\begin{equation}
    \label{eq:trivHam}
 H_{\mbox{\small{eff}}}^0 = \frac{\delta E_0}{\delta n(\vec p, \vec
r)} =
\frac{(\vec p +
e \vec A(\vec r))^2}{2m_b}
\end{equation}
and yields the conductivity $\tilde \sigma$ as given in Eq.
(\ref{eq:sigmatilde}) in the $\tau \rightarrow \infty$ limit as
usual.  Now, if we try to na\"{\i}vely account for the mass
renormalization by replacing the band mass $m_b$ by the effective
mass $m^*$ everywhere (as well as replacing $\Delta \omega_c$ by
$\Delta \omega_c^*$), we call the result the na\"{\i}ve
semiclassical conductivity $\tilde \sigma^{\mbox{\small n}}$.  As
mentioned before, this na\"{\i}ve approach violates the $f$-sum
rule and Kohn's theorem. None-the-less the na\"{\i}ve
conductivity will provide the starting point for our modified
semiclassical calculation.

We now
calculate the effective Hamiltonian (\ref{eq:Heff}) from
our energy functional (Eq.
(\ref{eq:FLHam})).  We find
\begin{equation}
    \label{eq:JHam}
     H_{\mbox{\small{eff}}}
 = \frac{(\vec p + e
\vec A(\vec r) - \frac{A_1 m_b}{n_e e}
\vec J(\vec r))^2}{2m^*} + {\cal O}( \vec J^2 )
\end{equation}
which to lowest order in  the perturbing electromagnetic field
is exactly the above trivial Hamiltonian (Eq.(\ref{eq:trivHam}))
 but with a
renormalized mass $m^*$  and a renormalized vector
potential
\begin{eqnarray}
    \label{eq:thisone}
    \vec A_{\mbox{\small{re}}} &=&  \vec A -
\frac{A_1 m_b}{n_e e^2} \vec J
 \\ &=&
        \vec A - \frac{(m^* - m_b)}{n_e e^2}    \vec J.
\end{eqnarray}
Equation (\ref{eq:thisone}) is equivalent to using
a renormalized electric field
\begin{eqnarray}
    \vec E_{\mbox{\small{re}}} &=& \vec E +
 \frac{(m^* - m_b)}{n_e e^2}
\frac{\partial   \vec J}{\partial t} \\
                    &=& \vec E -  i \omega
\frac{(m^* - m_b)}{n_e e^2}
 \vec J_{_.}
\end{eqnarray}
We can neglect the  associated magnetic field
 renormalization to first order.

Since the effective Hamiltonian $ H_{\mbox{\small{eff}}}$ looks
like the trivial Hamiltonian $ H_{\mbox{\small{eff}}}^0$, we can
calculate the current by using the na\"{\i}ve conductivity and
the renormalized electric field
\begin{eqnarray}
    \vec J &=&\tilde \sigma^{\mbox{\small n}}
    \vec E_{\mbox{\small{re}} } \\
            &=&\tilde \sigma^{\mbox{\small n}}
         \left[\vec E - i \omega \frac{(m^* - m_b)}{n_e e^2}
 \vec J \right]_.
\end{eqnarray}
We then  solve for the self-consistent current
\begin{equation}
    \vec J =  \left[ 1 -  i \omega \frac{(m_b - m^*)}{n_e e^2}
\tilde\sigma^{\mbox{\small n}} \right]^{-1}
\tilde\sigma^{\mbox{\small n}}
 \vec E
\end{equation}
and thus extract the conductivity $\tilde \sigma$
for the system of
quasiparticles of effective mass $m^*$ in a magnetic
field $\Delta B$,
\begin{equation}
  \tilde \sigma = \left[ 1 -  i \omega \frac{(m_b - m^*)}{n_e e^2}
\tilde\sigma^{\mbox{\small n}} \right]^{-1}
 \tilde\sigma^{\mbox{\small n}}
\end{equation}
In terms of the resistivity  $\tilde \rho$ (which is the inverse
of $\tilde \sigma$)  this can be written simply as
 \begin{equation}
    \label{eq:modsem}
    \tilde \rho = \tilde \rho^{\mbox{\small n}}
- \frac{i\omega (m_b-m^*)}{n_e e^2} 1.
\end{equation}
where $1$ is the identity matrix and $\tilde \rho^{\mbox{\small
n}} = (\tilde \sigma^{\mbox{\small n}})^{-1}$.  It should be
noted that as long as $\tilde \rho^{\mbox{\small n}}$ satisfies
the sum rule (\ref{eq:rhonlim}) then $\tilde \rho$ satisfies the
desired sum rule (\ref{eq:qpsum}).

The full prescription for calculating the resistivity (and hence
the response) of the fractional quantized Hall state  in this
modified semiclassical formalism is to calculate the na\"{\i}ve
quasiparticle conductivity $\tilde \sigma^{\mbox{\small n}}$
using Eq.  (\ref{eq:sigmatilde}) in the $\tau \rightarrow \infty$
limit, where we replace all occurrences of the cyclotron
frequency $\Delta \omega_c$ by the mass renormalized cyclotron
frequency $\Delta \omega_c^*$.  (The infinite sum in this
equation is performed explicitly in Appendix \ref{app:semisum}).
Next we invert to get the associated resistivity $\tilde
\rho^{\mbox{\small n}}= (\tilde \sigma^{\mbox{\small n}})^{-1}$.
We then add the diagonal effective mass correction term (Eq.
(\ref{eq:modsem})) to get the quasiparticle resistivity and the
off-diagonal Chern-Simons correction  term (Eq.
(\ref{eq:SemRPA2}))  to get the true resistivity $\rho$.
Altogether
\begin{equation}
    \label{eq:rhocomplete}
    \rho =  \tilde \rho^{\mbox{\small n}} -
 \frac{i\omega (m_b-m^*)}{n_e e^2}
\left[ \begin{array}{cc} 1 & 0  \\
                          0 & 1 \end{array} \right]
 + \frac{2\pi \hbar \tilde \phi} { e^2   }
                \left[ \begin{array}{cc} 0 & 1  \\
                          -1 & 0 \end{array} \right]_.
\end{equation}
The resistivity $\rho$
can then be converted to an electromagnetic response $K$ using
Eq. (\ref{eq:SemRPA1}).

\subsection{Modified RPA}
\label{sub:modRPA}

The above semiclassical prescription (\ref{eq:modsem}), that one
accounts for mass renormalization by adding a constant
resistivity, looks very much like the RPA prescription
(\ref{eq:SemRPA2}) for taking into account the effect of the
Chern-Simons field by simply adding a constant to the resistivity
tensor.  This encourages us to try to account for the mass
renormalization in the RPA calculation by the following analogous
method.  We write a quantum mechanical energy functional  Eq.
(\ref{eq:FLHam}) where we now think of the phase space
distribution $n(\vec r, \vec p)$ as its quantum mechanical
analog, the Wigner function.  As above, we can differentiate to
get the effective single particle Hamiltonian (\ref{eq:Heff})
except now we should think of this as a quantum mechanical
operator.   As above in Eq. (\ref{eq:trivHam}), if we neglect the
mass renormalization by setting the Fermi liquid coefficient
$A_1$ to zero, our effective Hamiltonian is the same trivial free
particle Hamiltonian except that now, $\vec p$ and $\vec r$ must
be treated as quantum mechanical operators.  Using this as a
single particle Hamiltonian, we know how to calculate the
electromagnetic response $K^0$ given in Eq.  (\ref{eq:K0def})
which we write in terms of the quantum mechanical conductivity
matrix $\tilde s$ in Eq. (\ref{eq:sfree}).
Again, if we
na\"{\i}vely substitute the effective mass $m^*$ and the
mass-renormalized cyclotron frequency $\Delta \omega_c^*$ for the
band mass $m_b$ and the cyclotron frequency $\Delta \omega_c$ in
Eqs. (\ref{eq:sfree})-(\ref{eq:Sigmadef}) we obtain the na\"{\i}ve
quantum mechanical
conductivity $\tilde s^{\mbox{\small n}}$.  Once again, we know
that this expression violates the $f$-sum rule and Kohn's
theorem.

    Now if we add in the interaction term in the energy
functional, we get the effective Hamiltonian  Eq.
(\ref{eq:JHam}) except that now $\vec p$ and $\vec r$ are
operators, and $\vec J$ is a current expectation value.  Again,
this Hamiltonian is just the trivial Hamiltonian with a
renormalized mass and a renormalized vector potential.  We can
thus follow the rest of the modified semiclassical prescription
exactly.

Thus, the complete prescription for the modified RPA is to first
calculate the na\"{\i}ve quantum mechanical conductivity $\tilde
s^{\mbox{\small n}}$ using Eqs.
(\ref{eq:sfree})-(\ref{eq:Sigmadef}) where we substitute  the
mass-renormalized cyclotron frequency $\Delta \omega_c^*$ for
the cyclotron frequency $\Delta \omega_c$.  The sums that occur
in this equation are simplified in Appendix \ref{app:RPAsum}.
We then invert this conductivity matrix to get the na\"{\i}ve
resistivity $\tilde \rho^{\mbox{\small n}}= (\tilde
s^{\mbox{\small n}})^{-1} $.  Finally we include the
mass-renormalization and Chern-Simons terms exactly as we did
for the modified semiclassical case by using Eq.
(\ref{eq:rhocomplete}) to get the resistivity $\rho$.  Again the
resistivity can  be converted to an electromagnetic response
using Eq.  (\ref{eq:SemRPA1}).

Although this Fermi-liquid approach is certainly appropriate in
the semiclassical regime, it is not as clear that it is
appropriate for correcting our RPA calculation.  Formally one
should probably use a diagrammatic expansion in the
electron-electron interaction to calculate both the value of the
effective mass $m^*$, and the correction to the conductivity.
The problem with this approach is that as shown by
HLR\cite{Tome}, these types of calculations are plagued with
divergences (although the coefficients of the diverging terms may
be small so that they are easier to ignore in practice).
Furthermore, since the mass is so greatly renormalized, such a
perturbative calculation might converge only very slowly.
None-the-less, we believe that our approach is at least
reasonably accurate as well as being the simplest approach that
still satisfies all of the sum rules.  Furthermore, it should be
noted that the form of our approximation (\ref{eq:modsem}) for
the quasiparticle resistivity $\tilde \rho$ coincides with an
approximation proposed by Ando in 1976 for the total resistivity
of a two-dimensional electron system\cite{Ando}.  Ando derived
this approximation using a particular short-range form of the
electron-electron interaction which he treated in lowest order
perturbation theory, corresponding to a Hartree-Fock
approximation for the electron self energy and a ladder
approximation for the vertex correction to the polarization
bubble.  Although Ando also considers the possible effects of
impurity scattering, his analysis is restricted to the $q=0$
limit, and of course he does not include a Chern-Simons
contribution in  his model.

\section{Numerical Results}
\label{sec:res}

We begin by  limiting  our attention to the series  of quantum
Hall states given by  $\nu = p/(2p+1)$.  This is the most stable
experimentally observed series of states, and is thus the most
interesting.  We  focus on these states by setting the flux
attached to each quasiparticle to be exactly two quanta ($\tilde
\phi = 2m =2$).  The results we would find for the more general
case $\nu = p/(2mp+1)$ are quantitatively very similar to the
results for $\nu = p/(2p+1)$ except the poles in the
density-density response function have much smaller weight in
general.

\subsection{Semiclassical}

 We first examine the case where there is no mass renormalization
$(m^*=m_b)$ and where we turn off the direct Coulomb interaction
by taking the limit of large dielectric constant ($\epsilon
\rightarrow \infty$), or by setting $V=0$ in Eq.
(\ref{eq:SemRPA1}).  In the semiclassical approximation, we have
used Eq. (\ref{eq:sigmatilde}) in the low scattering ($\tau
\rightarrow \infty$) limit to calculate the quasiparticle
conductivity $\tilde \sigma$.  An equivalent, but more convenient
form of this equation is given in Appendix \ref{app:semisum}.  We
then convert this quasiparticle conductivity to a response
function $K$ by using Eqs.  (\ref{eq:SemRPA1}) -
(\ref{eq:rtfromst}).

 We are most interested in the poles in the density-density
response function $K_{00}$.  A pole in $K_{00}$ with respect to
frequency indicates the existence of a collective mode, and the
weight of the pole indicates the strength of the coupling of this
mode to a fluctuation in density.  In  Fig.  \ref{fig:sm1v0} we
show the location (heavy solid) of these poles in $K_{00}$ as a
function of reduced frequency ($\omega/\Delta \omega_c$)  and
reduced wavevector ($X=qR_{\Delta} = 2qp/k_{\mbox{\tiny F}}$).
The width of the striped bands around the lines of poles indicates
$q^{-2}$ times the relative weight of the poles.  In accordance
with Kohn's theorem, we see that the cyclotron mode (the mode at
$\omega = (2p+1) \Delta \omega_c = \omega_c$) has all of the
weight at long wavelength and moreover that this weight scales as
$q^2$.  It appears as though  some of the lines of poles get very
thin and disappear at certain wavevectors.  What is actually
happening here is just that the residue of the line of poles has
become too small to see on the scale of the graph shown.

We have shown results for  filling fractions $\nu =\frac{1}{3}$,
$\frac{3}{7}$, and $\frac{10}{21}$, corresponding to effective
Landau level fillings of $p=1$, $3$, and $10$  where $m=1$.  By
examining the $p=10$ case we see that the semiclassical
approximation has a very simple large $p$, large $X/R$ limit
(where $R=\omega/\Delta \omega_c$ as usual).  In this limit we see
the series of crossing straight lines with equal slopes but with
opposite signs.  More precisely, we have
\begin{equation}
    \label{eq:lin1}
    X_{\mbox{pole}} \sim \pm \frac{\pi}{2}
 \frac{\omega}{\Delta \omega_c} +
\left( n+\frac{1}{4} \right) \pi.
\end{equation}
where $n$ is an integer.  This behavior can be derived most easily
by using the analytic form for $\tilde \sigma$ described in
Appendix \ref{app:semisum} and expanding the Bessel functions for
large argument using Eq. 9.2.1 of ref. \cite{Abrom}.  The above
linear relation then follows after converting $\tilde \sigma$ to a
response function.

We can gain some insight into the physics behind this linear
relation with the following heuristic argument.  In the present
approximation, we have ignored the Coulomb interaction $V$ in Eq.
(\ref{eq:SemRPA1}).  Therefore,  in order to have a pole in the
electromagnetic response $K$, we must have a zero of the
determinant of the resistivity $\rho$.  But $\rho = \tilde \rho +
\rho_{cs}$ (See Eq.(\ref{eq:SemRPA2})) can only have a zero if the
quasiparticle resistivity  $\tilde \rho$ is large enough to cancel
the Chern-Simons resistivity $\rho_{cs}$.  Since $\rho_{cs}$ is
large, we can only have this cancellation very near to a pole in
$\tilde \rho$  which occurs when there is a zero eigenvalue of the
quasiparticle conductivity $\tilde\sigma$.  Thus, the poles in the
electromagnetic response $K$ occur at approximately the same
wavevector as the zero eigenvalues of the quasiparticle
conductivity tensor $\tilde\sigma$.

We now imagine a semiclassical  quasiparticle orbiting in the
magnetic field $\Delta B$ with a semiclassical cyclotron radius of
$R_{\Delta}$ in the presence of a perturbing electric field $\vec
E_{\mbox{\small eff}}$ which is the sum of the Chern-Simons and
actual electric fields. If the perturbation is applied at $\omega
\approx 0$ and at a wavevector $\vec q$ such that the wavelength
is less than the cyclotron radius ($2\pi/q << R_{\Delta} $), then
during the course of one orbit the quasiparticle experiences $\vec
E_{\mbox{\small eff}}$ in oppposite directions as shown in Fig.
\ref{fig:orbit1}.  When the quasiparticle is moving essentially
parallel to the wavevector $\vec q$ (the $\hatn x$ direction in
our convention) as shown by the dotted lines of the orbit in Fig.
\ref{fig:orbit1}, it experiences a quickly oscillating field.
Thus  there is no net energy loss or gain from moving through this
part of the orbit.  On the other hand, when the quasiparticle is
moving perpendicular to the wavevector (the $\hatn y$ direction)
as shown by the solid lines in Fig. \ref{fig:orbit1}, it
experiences the same force for an extended period of time.  This
is the part of the orbit where the quasiparticle can gain or lose
energy due to its interaction with the field.

We now use the fact that the semiclassical quasiparticle
conductivity
is just the Fourier
transform of the velocity autocorrelation function\cite{Pines}:
\begin{equation}
\label{eq:sigint}
         \tilde       \sigma_{ij} \propto \int_0^{\infty} d(t-t')
     e^{-i\omega (t-t')} \left< v_i(t) v_j(t') e^{i \vec q
\cdot [\vec r(t) - \vec r(t')]} \right>.
\end{equation}
Since the quasiparticle is undergoing cyclotron motion, $\exp\{i
\vec q \cdot [\vec r(t) - \vec r(t')]\}$ oscillates very quickly
when the particle is moving in the $\hatn x$ direction, and stays
constant when the particle is moving in the $\hatn y$ direction
(see Fig. \ref{fig:orbit1}).  Thus, when we integrate to obtain
the low frequency conductivity, we expect that $\tilde
\sigma_{yy}$ will, in general, be the largest component of the
conductivity tensor.  This assertion is verified by examining Eqs.
(\ref{eq:xxend}), (\ref{eq:xyend}), and (\ref{eq:yyend}) of
Appendix \ref{app:semisum} where we see that $\tilde \sigma_{yy}$
is of higest order in $q$, and thus dominates in the large $q$
limit.  We conclude that the zero eigenvalue of the quasiparticle
conductivity tensor (and hence the pole in the electromagnetic
response) must occur very close to the point where $\tilde
\sigma_{yy}=0$.

We now set $t'$ to be the time when the quasiparticle is at the
extreme right of its orbit such that $v_y(t')$ is large and
negative.  Clearly the exponential factor $\exp\{i \vec q \cdot
[\vec r(t) - \vec r(t')]\}$ is unity whenever the quasiparticle
returns to the extreme right of the orbit.  Furthermore,  if the
diameter of the orbit is approximately an  integer number of
wavelengths, then the exponential factor is approximately unity
when the quasiparticle is at the extreme left of the orbit also.
Now since $v_y(t)$ oscillates (with frequency $\Delta \omega_c$)
and is a maximum at the far left and a minimum at the far right,
we see that these two pieces will approximately cancel in the
integral in Eq. (\ref{eq:sigint}).   Thus if the diameter of the
orbit is approximately an  integer number of wavelengths, $\tilde
\sigma_{yy}$ will be zero, and hence a  there will be a pole in
the response function.  More careful analysis shows that the
condition for having a pole in the response function at zero
frequency is
\begin{equation}
2  R_{\Delta} = (n +
\frac{1}{4})\lambda =  \frac{2 \pi }{q} (n +
\frac{1}{4})
\end{equation}
 which is exactly the $\omega \rightarrow 0$ limit of Eq.
(\ref{eq:lin1}).  The ``$+\frac{1}{4}$'' is included because the
average separation of the two transverse parts of the orbit is
somewhat less than the the full diameter of the orbit (See Fig.
\ref{fig:orbit1}).  These poles at zero frequency were first
predicted by HLR\cite{Tome}, and are somewhat analogous to the
``geometric resonances'' found in the propagation of acoustic
waves in a direction perpendicular to an applied magnetic field in
three-dimensional metals\cite{Harrison}.

Now we consider the effect of nonzero frequency.  When the wave is
in motion, we want to arrange that the phase of the wave when the
quasiparticle is at one transverse part of the orbit is the same
as the phase of the wave when the quasiparticle reaches the other
transverse part of the orbit such that  $\exp\{i \vec q \cdot
[\vec r(t) - \vec r(t')] - i\omega t\}$ is equal at the extreme
left and extreme right of the orbit.  This is most easily
visualized by considering only the coordinate of the quasiparticle
which is parallel to the wavevector (the $\hatn x$ coordinate in
our previous convention).   Now consider the linear world lines of
the crests of the wave and the sinusoidal world line of the
quasiparticle as shown in Fig.  \ref{fig:world}.  There are two
possible ways to have exactly no net contibution to the integral
(\ref{eq:sigint}).  The first possibility is that every time the
particle moves to the right it begins and ends at the same phase
of the wave (Case I in Fig.  \ref{fig:world}).  Alternately, the
particle can begin and end at the same phase of the wave every
time it moves to the left (Case II in  Fig.  \ref{fig:world}).
Note that these two cases are not equivalent since the phase of
the wave is different at the beginning of each orbit.   It is not
too hard to see that these two possible conditions are exactly the
conditions written above in Eq.  (\ref{eq:lin1}).  If either of
these conditions are met, then the corresponding contributions to
the integral (\ref{eq:sigint}) from the extreme left and extreme
right of the orbit cancel, and we should have a zero of $\tilde
\sigma_{yy}$ and hence a pole in the response.

Finally we consider the special case when $\omega = n \Delta
\omega_c$.  Since $\omega$ and $\Delta \omega_c$ are commensurate,
we can have energy absorbed and re-emitted at the applied
frequency, and hence a pole in the quasiparticle conductivity.  It
is easy to see that in this conditon ($\omega = n \Delta
\omega_c$) both above Cases I and II (both signs of Eq.
(\ref{eq:lin1})) are satisfied simultaneously.  This would
corresponds to the ``crossing'' of the lines of poles in the above
spectrum (Fig.  \ref{fig:sm1v0}) at multiples of the effective
cyclotron frequency $\Delta \omega_c$ as predicted by Eq.
(\ref{eq:lin1}).  Note, however, that the lines of poles  in Fig.
\ref{fig:sm1v0} do not actually cross at these points.  The fact
that the quasiparticle conductivity has a pole rather than a zero
at these special frequencies creates  a ``level-repulsion''
keeping the lines of poles from crossing.  In terms  of the
integral  (\ref{eq:sigint}), the pole in the conductivity occurs
because the phase of $\exp\{i \vec q \cdot [\vec r(t) - \vec
r(t')] - i\omega t\}$ is the same at the beginning of each orbit.
Thus, any small noncancellation of the contributions to the
integral will occur identically for each orbit, and thus these
terms will add and cause a diverging conductivity.

Although the effects of this semiclassical orbiting behavior are
most obvious in the large $X$ and large $p$ limits, the same
general behavior is seen even for $p=1$ (although the validity of
the semiclassical approximation at low $p$ and low frequency is
questionable).  One notes that the semiclassical theory predicts a
series of magnetoroton-like minima, ie minima in the dispersion
curve of the magnetoexciton (the lowest neutral excited mode).
The first of these minima  occurs for any given $p$ approximately
where the first zero frequency mode would occur in the large $p$
limit ($X \approx \frac{5}{4} \pi$).

\subsection{Unrenormalized RPA}

We now consider the RPA in the same limit where the mass is
unrenormalized ($m^* = m_b$).  This is the calculation considered
by Lopez and Fradkin in the $q \rightarrow 0$ limit\cite{Lopez2}.
However, unlike Lopez and Fradkin, we will  begin by  considering
the case when the Coulomb interaction is turned off ($\epsilon
\rightarrow \infty$). In this case we calculate the mean-field
unperturbed response function $K^0$ using Eqs.  (\ref{eq:K0def})-
(\ref{eq:Sigmadef}) which are given as a simplified single sum in
Appendix \ref{app:RPAsum}.  This mean-field result is then
converted into the full RPA response function using the RPA
equation (\ref{eq:RPA}).  The results of such a calculation are
shown in Fig.  \ref{fig:rm1v0}.  Once again we have shown the
location of the pole (heavy solid) in the density-density response
function $K_{00}$ as a function of reduced wavevector and reduced
frequency, and the width of the striped bands indicate $q^{-2}$
times the relative weights of the poles.  As before we have shown
results for filling fractions $\nu =\frac{1}{3}$, $\frac{3}{7}$,
and $\frac{10}{21}$ corresponding to effective Landau level
fillings of $p=1$, $p=3$, and $p=10$ respectively  where $m=1$.

 In the large $p$ limit, the observed behavior looks very much
like the above semiclassical picture of particles undergoing
cyclotron-like oscillations while interacting with the
Chern-Simons force.  This is to be expected since the
semiclassical picture is thought to be accurate in this limit.  We
also note that in all cases except $p=1$, we  have additional
modes of small residue in the RPA calculation that did not show up
in the semiclassical case. The difference between the RPA and the
semiclassical calculations is that for the semiclassical
quasiparticle conductivity, the residue at the pole $n\Delta
\omega_c$ is proportional to projection matrix $[V^n_i]^* V^n_j$
(see Eq.  \ref{eq:sigmatilde}).  Such projection matrices have a
zero eigenvalue.  This singular situation causes the disappearance
of additional solutions of the equation for the poles of the
system.  Similarly, for the $p=1$ case of the RPA when we
calculate $K^0$ (see Appendix \ref{app:RPAsum}) there is only one
way for a virtual transition to take place with a given energy
difference  $n \Delta \omega_c^*$ (a jump from the single filled
level to the $n^{th}$ empty level and back).  When only a single
process contributes to a given pole, then the residue takes the
projection-like form $<i| j_i|f><f|j_j|i>$ where $|i>$ is the
initial state and $|f>$ is the final state.  In the $p>1$ case for
the RPA several processes can contribute to each pole, except the
lowest one, and thus interfere to ruin the perfect projection form
of the residues in the  quasiparticle response function, and so we
see additional poles.  However, to the extent that the
semiclassical calculation approximates the RPA, we expect that the
additional poles will have a very small residue.

If we were to add a Coulomb interaction by taking $\epsilon$
finite (for typical experimental parameters the density of
electrons is $n_e = 10^{-11} \mbox{cm}^{-2}$ and the dielectric
constant $\epsilon = 12.6$) the resulting response function would
be quite similar to Fig.  \ref{fig:rm1v0}, so we have not shown it
here.  The main effect of the Coulomb interaction is to push the
weight of the poles up to higher frequency modes.  In addition,
very small shifts in the frequency of the modes are also seen.
Since the Coulomb interaction couples to density fluctuations, it
is  mainly those modes corresponding to large residues of the
density-density response function that are affected.  At small
$q$, the cyclotron mode has all the weight, so it is shifted the
most.  However, exactly at $q=0$, Kohn's theorem must be satisfied
so the total weight of the cyclotron  pole  (which is the sum of
two modes in the $p=1$ case) must stay the same.  The Coulomb
interaction also seems to slightly reduce the depth of the
magnetoexciton minima.

\subsection{Modified RPA}
\label{sub:modres}
Here we consider the effect of using the renormalized mass.  The
method of calculating the response function -- which we outline
here -- was described in section \ref{sub:modRPA}.  We begin by
using Eq.  (\ref{eq:sfree}) to calculate
 the na\"{\i}ve
quasiparticle conductivity $\tilde s^{\mbox{\small n}}$ in a field
$\Delta B$  where here we have replaced $\Delta \omega_c$ by
$\Delta \omega_c^*$ in Eqs.  (\ref{eq:sfree})-(\ref{eq:Sigmadef}).
  Again, the
necessary sums are given in a simplified form in Appendix
\ref{app:RPAsum}.   We then set $\tilde \rho^{\mbox{\small n}}=
(\tilde s^{\mbox{\small n}})^{-1} $ and use Eq.
(\ref{eq:rhocomplete}) to include the mass-renormalization and
Chern-Simons terms  to yield the resistivity $\rho$.  Finally, the
resistivity is converted to an electromagnetic response $K$ using
Eq.  (\ref{eq:SemRPA1}).   Results of these calculations are shown
in Fig.  \ref{fig:rm3vfinal}.

 Figure \ref{fig:rm3vfinal}  should be compared to Fig.
\ref{fig:rm1v0} where we have not renormalized the mass or
included the Coulomb interaction.  First, however, we note that if
we were to make the na\"{\i}ve RPA approximation (simply inserting
$m^*$ into the Hamiltonian in place of $m_b$ to calculate $\tilde
\rho^{\mbox{\small n}}$, then including the Chern-Simons term and
converting this  into a response via $K^{-1} = T^{-1}[\tilde
\rho^{\mbox{\small n}}+ \rho_{cs}] T^{-1}$ ) the end result would
be exactly like Fig.  \ref{fig:rm1v0} except that the vertical
scale would now be $\omega/\Delta \omega_c^*$ as it is in Fig.
\ref{fig:rm3vfinal} and all the weights would be scaled by the
same factor.    Although this approach gives the correct energy
scale for the low energy excitations, it fails to satisfy the
$f$-sum rule, and the cyclotron frequency is incorrect.

On the other hand, if we include the mass renormalization properly
by using the modified RPA (Eqs.  (\ref{eq:sfree}),
(\ref{eq:Sigmadef})
(\ref{eq:rhocomplete}), (\ref{eq:SemRPA1}) as described above) but
still ignore the Coulomb interaction, we find that the cyclotron
frequency is pushed up to its correct value ($\omega_c = m^*
(2p+1) \Delta \omega_c^*/m_b$) and the weights of the poles
satisfy the $f$-sum rule.  At the same time, the low lying modes
in such a calculation are virtually unchanged from the above
mentioned Fig.  \ref{fig:rm1v0} rescaled.

Finally, when we include the Coulomb interaction as well as the
mass renormalization in the modified RPA, we obtain Fig.
\ref{fig:rm3vfinal}.  Here we show the results of such a
calculation for filling fractions $\nu = \frac{1}{3},\frac{3}{7}$,
and $\frac{10}{21}$ corresponding to effective Landau-level
fillings $p=1,3$, and $10$ respectively  where $m=1$.  Once again
the width of the striped bands indicates $q^{-2}$ times the weight
of the pole (solid). We have used the experimentally relevant
parameters $n_e = 10^{-11} \mbox{cm}^{-2}$, $\epsilon = 12.6$, and
an effective mass of $m^* = 3.9m_b$ for illustrative purposes. As
discussed in Sec. \ref{sec:mstar}, the value $m^* = 3.9m_b$ is
thought to be approximately correct for certain relevant
experimental conditions.  ($m^* = 4m_b$ was avoided simply because
there is no reason to assume that the effective mass should be an
integer multiple of the band mass).   Although, as we discussed in
Sec.  \ref{sec:mstar} , the effective mass is in general a
function of the magnetic field, we will treat it as a constant
here.  The  results given in Fig.  \ref{fig:rm3vfinal} are the
complete theory including Coulomb interaction and the mass
renormalization due to the Fermi liquid interaction $A_1$.

 We note that the general structure of Fig.  \ref{fig:rm1v0} and
Fig. \ref{fig:rm3vfinal} are similar, except that the cyclotron
frequency has been pushed up from the effective cyclotron
frequency $\omega_c^* = (2p+1) \Delta \omega_c^*$ in Fig.
\ref{fig:rm1v0} (where we read the vertical scale as
$\omega/\Delta \omega_c^*$)  to the bare cyclotron frequency
$\omega_c = m^* (2p+1) \Delta \omega_c^*/m_b$ in Fig.
\ref{fig:rm3vfinal}.  In the cases of $p=3$ and $p=10$, the
cyclotron frequency is pushed off of the top of the graph shown.
As mentioned above, Fig.  \ref{fig:rm1v0} rescaled provides a good
approximation of the  low lying excitations when the Coulomb
interaction is ignored.  By comparing Fig.  \ref{fig:rm1v0} and
Fig. \ref{fig:rm3vfinal}, we see that the inclusion of the direct
Coulomb interaction causes a significant change in the shape of
the lowest magnetoexciton curve.  In particular the magnetoexciton
minima are much less pronounced.  The Coulomb interaction has a
much larger effect in the modified RPA than it did in the simple
RPA primarily because the overall energy scale is smaller.

In the case of $p=1$ ($\nu=\frac{1}{3}$), these results may be
compared directly with results of the single mode
approximation\cite{Girvin1} which is believed to be quite accurate
near the magnetoroton minimum. One finds that the actual minimum
is significantly deeper than that found in either the RPA, the
modified RPA, or the semiclassical approximation.  A relatively
deep magnetoroton minimum has also been found in numerical work on
finite systems\cite{SongHe,Morf} and in the analytic approach of
Zhang, Hanson, and Kivelson\cite{Zhang}.

 We speculate that the inclusion of Coulomb ladder diagrams (ie,
the attraction between the quasihole and quasiparticle of the
exciton) would enhance the size of the magnetoroton minimum
relative to that found in the RPA or modified RPA, and would
perhaps bring the perturbative Chern-Simons calculation into
better agreement with the other calculations in this regime.  We
also speculate that the Coulomb ladder diagrams may be relatively
less important in the case of large $p$, where the charges of the
quasiparticle and quasihole are small, so that the modified RPA
may give an accurate description of the dispersion of the lowest
excitation mode in this case.  We find that for very large $p$,
the dispersion curves show a series of deep minima which are
equally well respresented in the RPA, modified RPA, or
semiclassical approximations.  However, at $p=10$ ($\nu =10/21$)
there are still significant differences in the size of the exciton
minima according to Figure \ref{fig:sm1v0}, \ref{fig:rm1v0}, and
\ref{fig:rm3vfinal}.  Again we note that the depths of the
magnetoexciton minima are  somewhat suppressed by the Coulomb
interaction in the modified RPA.  The other main contribution of
the Coulomb interaction is simply to push the weight of the poles
to higher frequency modes.

\section{Summary}
\label{sec:summ}

In this paper we have reviewed the Chern-Simons construction that
allows us to think of certain fractional quantized Hall states as
integer quantized Hall states of fermionic quasiparticles bound to
an even number of flux quanta.  The electromagnetic response
function was first calculated in a semiclassical approximation and
within the RPA.  If one uses the bare electron band mass $m_b$ in
these calculations, one obtains an incorrect frequency scale for the
low energy excitations; if one simply replaces $m_b$ by an effective
mass $m^*$ which is chosen to give the correct scale for the low
energy excitations, then one obtains an incorrect value for the
cyclotron energy, in violation of Kohn's theorem, and one obtains
intensities that fail to satisfy the $f$-sum rule.  A modified RPA
was then constructed that  accounts for the effective mass
renormalization by using a Fermi liquid theory approach.   The
results of the modified RPA calculation properly satisfy the $f$-sum
rule and Kohn's theorem and also have the low energy excitation
spectrum approximately correct.  In the semiclassical regime ($\nu$
approaching an even denominator fraction and large wavelength
compared to the magnetic length) we clearly see the orbiting
behavior that results in geometric resonances including a series of
magnetoexciton minima at increasing wavevector.  At  $\nu =
\frac{1}{3}$, the magnetoroton minimum is not as deep in our
approximation as previous works predict.  (We speculate that Coulomb
ladder diagrams which have not been included within the RPA may
increase the depth.)  Finally, we note  that within the RPA the main
effect of the direct inter-quasiparticle Coulomb interaction is to
slightly reduce the depth of the magnetoexciton minima and to push
some of the weight of the poles of the response function up to modes
of higher frequencies.

All of the approximations discussed in this paper omit the possible
effects of quaiparticle scattering.  When such effects are taken
into account, we expect that in general; the higher excitation
modes will acquire a finite energy width, as they can generally
decay into two or more modes of lower energy, while conserving
momentum and energy.  If the decay rate becomes larger than the
spacing between modes for some regions of the $k,\omega$ plane, then
the energy spectra indicated in Fig. \ref{fig:rm3vfinal} will cease
to be meaningful in that region.  By contrast, we expect that the
lowest energy branch will remain perfectly sharp, in the absence of
impurity scattering, at least near the magnetoexciton minima because
there are no lower lying excitations to decay to with conservation
of energy and momentum.

\vspace{10pt}

{\centerline{\small
    ACKNOWLEDGEMENTS}}

The authors gratefully acknowledge helpful discussions with Ady
Stern, Song He, Rudolf Morf, and Onuttom Narayan.  We would also
like to thank Song He for making his simulation data available to
us prior to publication.   This work was supported by the
National Science Foundation Grant DMR-91-15491.  One of the
authors (SHS) also acknowledges support from a National Science
Foundation Fellowship.

\appendix
\setcounter{section}{0}
\renewcommand{\theequation}{\Alph{section}.\arabic{equation}}
\setcounter{equation}{0}

\section{Quantum Mechanical Sum}
\label{app:RPAsum}

To reduce the sums in the definition of
 the $\Sigma_j$ (Eq. (\ref{eq:Sigmadef})) to single sums, we start
by reparametrizing our dummy variables using $n=m-l$ such that
\begin{eqnarray}
 \Sigma_j &=&       \frac{e^{-Y}}{p} \sum_{n=1}^{\infty}
  \frac{ n Y^{n-1} }{R^2 -
n^2} S_j(n,Y)\\
S_j(n,Y) &=& \sum_{l=\max(0,p-n)}^{p-1} \!\!\!\!\!\! G_j(n,l,Y)
\end{eqnarray}
where
\begin{eqnarray}
    G_j(n,l,Y) &=&  \\
\nonumber  \frac{l!}{(n+l)!}&& \left[  L_l^n(Y) \right]^{2-j}
\left[(n-Y)L_l^n(Y) + 2Y\frac{dL_l^n(Y)}{dY} \right]^j,
\end{eqnarray}
and $L^n_l(Y)$ is the associated Laguerre polynomial.  In the
calculation of $\tilde s$ to get $K^0$  for the RPA in section
\ref{sub:K0}, we use $R=\frac{\omega}{\Delta\omega_c}$ whereas in
the calculation of $\tilde s^{\mbox{\small n}}$ for  the modified
RPA we use $R=\frac{\omega}{\Delta\omega_c^*}$.  In this
notation, the $(n,l)^{th}$ term in the $S_j$  sum represents a
particle making a virtual transition from the $l^{th}$
Landau-level up $n$ levels and back.  This form for $\Sigma_j$ is
also quite appealing physically since it groups terms of the sum
in terms of which pole they contribute to.  It will be convenient
to think of the sum over $l$ as the difference of two sums both
of whose lower limit is zero.  In other words
\begin{equation}
S_j(n,Y) =
\left[ \sum_{l=0}^{p-1}
\!G_j(n,l,Y) \right]
 - \theta(p-n)\!\!\left[\!\sum_{l=0}^{p-n-1}\!\!\!\!
 G_j(n,l,Y)\right]
\end{equation}
where $\theta$ is the step function
\begin{equation}
 \theta(x) =  \left\{ \begin{array}{cc} 0& \hspace{20pt}
        x\le 0 \\
           1 & \hspace{20pt} x>0 . \end{array} \right.
\end{equation}
The point of this  appendix
is to perform the sums over $l$ to yield
 closed form expressions for $S_j(n,Y)$.

To perform our sums it will be necessary to evaluate
 the three quantities
\begin{eqnarray}
T_n^\alpha(Y) &=& \sum_{m=0}^n \frac{m!}{(m+\alpha)!}
 \left[ L_m^\alpha(Y) \right]^2 \\
U_n^\alpha(Y) &=& \sum_{m=0}^n \frac{m!}{(m+\alpha)!} L_m^\alpha(Y)
\frac{d}{dY} L_m^\alpha(Y) \\
V_n^\alpha(Y) &=& \sum_{m=0}^n \frac{m!}{(m+\alpha)!} \left[
\frac{dL_m^\alpha(Y)}{dY} \right]^2
\end{eqnarray}
These can be calculated by first considering the
Christoffel-Darboux
formula\cite{Grad}
\begin{eqnarray}
\label{eq:CDsum}
\sum_{m=0}^n &\frac{m!}{(m+\alpha)!}& L_m^\alpha(x) L_m^\alpha(y) =\\
\nonumber
&\frac{(n+1)!}{(n+\alpha)! (x-y)}& \left[L_n^\alpha(x)
 L_{n+1}^\alpha(y) -
L_{n+1}^\alpha(x) L_n^\alpha(y) \right].
\end{eqnarray}
By setting $x=Y+\delta$ and $y=Y$ and differentiating with
respect to $\delta$ at
$\delta = 0$, and then using the identity\cite{Grad}
\begin{equation}
    \label{eq:dL}
    \frac{dL_n^\alpha(Y)}{dY} = -L_{n-1}^{\alpha+1}(Y)
\end{equation}
where a Laguerre polynomial of negative lower index is
defined here to be zero,
one easily derives
\begin{equation}
    \label{eq:sum1}
T_n^\alpha(Y) =
\frac{(n+1)!}{(n+\alpha)!} \left[L_n^\alpha(Y)
 L_{n}^{\alpha+1}(Y) -
L_{n-1}^{\alpha+1}(Y) L_{n+1}^\alpha(Y) \right].
\end{equation}
Differentiating this result yields
\begin{equation}
    \label{eq:sum2}
 U_n^\alpha(Y) = \frac{1}{2}
\frac{(n+1)!}{(n+\alpha)!} \left[L_{n-2}^{\alpha+2}(Y)
L_{n+1}^{\alpha}(Y) -
L_{n}^{\alpha}(Y) L_{n-1}^{\alpha+2}(Y) \right].
\end{equation}
The easiest way to find $V_n^\alpha(Y)$  without running into
divisions by zero is to write the derivatives as limits
\begin{eqnarray}
 V_n^\alpha(Y)&=&\lim_{\gamma,\epsilon,\delta  \nonumber
\rightarrow 0} \sum_{m=0}^n \frac{m!}{(m+\alpha)!} \left[
\frac{L_m^\alpha(Y+\gamma+\delta) -L_m(Y+\gamma)}{\delta} \right]
\\  &\times&   \hspace{20pt}
\left[ \frac{L_m^\alpha(Y+\epsilon) - L_m^\alpha(Y)}{\epsilon}
\right]
\end{eqnarray}
 such that we end up with four terms in the
Christoffel-Darboux form.  The sums are then performed using Eq.
(\ref{eq:CDsum}). At this point one must  be very careful in
taking the limits.  The easiest way to do this is to expand each
resulting Laguerre polynomial in a Taylor series around $Y$ to
{\it third order} such that the parameters $\gamma,\epsilon$ and
$\delta$ no longer occur inside the arguments of the polynomials.
(At the end of the calculation it is easy to see that higher
order terms are irrelevant since we will take the small
parameters $\gamma,\epsilon$ and $\delta$ to zero anyway).
Finally one can take the limits and find that the first and
second order terms of the expansion vanish leaving the result
$$
    V_n^\alpha(Y) =  \frac{(n+1)!}{(n+\alpha)!} \left[ \frac{1}{6}
\left\{L_{n-2}^{\alpha+3}(Y)
L_{n}^{\alpha}(Y) -
L_{n-3}^{\alpha+3}(Y) L_{n+1}^{\alpha}(Y) \right\} \right.
$$
\begin{equation}
    \label{eq:sum3}
 \hspace{10pt} \left. + \frac{1}{2} \left\{
L_{n-1}^{\alpha+2}(Y)
L_{n-1}^{\alpha+1}(Y) -
L_{n-2}^{\alpha+2}(Y) L_{n}^{\alpha+1}(Y) \right\} \right]
\end{equation}
where we have used the above Laguerre polynomial
 identity (\ref{eq:dL})
several times.

We can now use the three
derived sums ($T,U$ and $V$) to perform the $l$
sums in $S_j(n,Y)$
Using Eq. (\ref{eq:sum1}) and the
definition of $T_\alpha^n(Y)$ we have immediately
\begin{equation}
    S_0(n,Y) = T_{p-1}^n(Y)-\theta(p-n)T_{p-n-1}^{n}(Y)
\end{equation}
and similarly we use the sums (\ref{eq:sum1})
 and (\ref{eq:sum2}) to yield
\begin{eqnarray}
    S_1(n,Y) = \left[(n-X)T_{p-1}^n(Y)+2YU_{p-1}^n(Y)\right]  \\
    \nonumber
 \hspace{10pt}  -\theta(p-n)\left[(n-Y)T_{p-n-1}^n(Y)
 +2YU_{p-n-1}^n(Y)\right]
\end{eqnarray}
Now by using the Laguerre polynomial
identities 8.97.4 and 8.97.5 from
 ref. \cite{Grad}
this can be reduced to
\begin{eqnarray}
S_1(n,Y) = \frac{p!}{(p+n-1)!} L_{p-1}^{n+1}(Y) L_p^{n-1}(Y) \\
\nonumber
 -\theta(p-n)\frac{(p-n)!}{(p-1)!}L_{p-n-1}^{n+1}(Y)
 L_{p-n}^{n-1}(Y)
\end{eqnarray}
And finally we have
\begin{eqnarray}  \nonumber
S_2&(&n,Y) = \left[(n-Y)^2T_{p-1}^n(Y) +4Y(n-Y)U_{p-1}^n(Y) +
\right.  \\
&&\left. \nonumber
4Y^2V_{p-1}^n(Y)\right]-\theta(p-n)\left[(n-Y)^2T_{p-n-1}^n(Y) +
\right.
\\
&&\left. 4 Y(n-Y)U_{p-n-1}^n(Y) +
4Y^2V_{p-n-1}^n(Y)\right].
\end{eqnarray}
Although these results look somewhat messy,
 they eliminate one infinite
sum which is beneficial for both analytic and numerical work.
\section{Semiclassical Sum}
\label{app:semisum}

When we take the $\tau \rightarrow 0$ limit
 we can rewrite the quasiparticle
conductivity given in Eq. (\ref{eq:sigmatilde})  as
\begin{equation}
    \tilde\sigma_{ij} = \frac{ipe^2}{\pi \hbar}
\sum_{n=-\infty}^{\infty}
\frac{V_i^{(n)} V_j^{(n)*}}{R-n}
\end{equation}
where $p$ is the number of effective Landau-levels filled which
need not be an integer. We use  $R=\frac{\omega}{\Delta \omega_c}$
for the semiclassical calculation and  $R=\frac{\omega}{\Delta
\omega_c^*}$ for the modified semiclassical approximation (which
then yields $\tilde\sigma^{\mbox{\small n}} $ as a result).

Using symmetry relations of the
Bessel functions (Eq. 9.1.5 from ref. \cite{Abrom}) the
antisymmetric quasiparticle conductivity matrix can be rewritten as
\begin{eqnarray}
    \tilde\sigma_{xx} &=& \frac{ipe^2}{\pi \hbar}
 \sum_{n=1}^{\infty}
\frac{n^2 J_n^2(X)}{X^2} \frac{2R}{R^2 - n^2}  \\
    \tilde\sigma_{xy} &=& \frac{pe^2}{\pi \hbar}
 \sum_{n=1}^{\infty}
\frac{n^2 J_n(X)J_n'(X)}{X} \frac{2}{R^2 - n^2}  \\
        \tilde\sigma_{yy} &=& \frac{ipe^2}{\pi \hbar}
\left[ \frac{[J_0'(X)]^2}{R} + \sum_{n=1}^{\infty}
[J_n'(X)]^2 \frac{2R}{R^2 - n^2}  \right]    .
\end{eqnarray}

In order to evaluate these sums,
 we first consider the general quantity
\begin{equation}
 W_\alpha = \sum_{n=1}^{\infty} \frac{J_{n+\alpha}(X)
 J_{n-\alpha}(X)}{R^2 - n^2} .
\end{equation}
Using the   integral identities 3.715.19 and 6.681.1 from ref.
\cite{Grad}
we can rewrite $W_\alpha$ as a sum over a double integral of
cosines.
\begin{eqnarray}  \nonumber
    W_\alpha= \frac{4}{\pi}
\sum_{n=1}^\infty \frac{(-1)^n}{R^2 - n^2}
    \int_0^{\frac{\pi}{2}} &dt&
 \int_0^{\frac{\pi}{2}} dt'\cos(2X\cos t \cos
t')\\
&\times& \cos(2\alpha t) \cos(2nt')
\end{eqnarray}
  The sum is
then performed by using sum 1.445.8 from ref \cite{Grad}
leaving  us with
\begin{equation}
W_\alpha = Q_\alpha+P_\alpha
\end{equation}
where
\begin{equation}
    Q_\alpha = \frac{-2}{\pi^2R^2}  \int_{0}^{\pi/2} \!\!\!\!\!  dt
\int_{0}^{\pi/2} \!\!\!\!\! dt' F_\alpha[t,t',X]
\end{equation}
\begin{equation}
    P_\alpha = \frac{2}{\pi R \sin(R \pi)}
\int_{0}^{\pi/2} \!\!\!\!\!  dt
\int_{0}^{\pi/2} \!\!\!\!\! dt' F_\alpha[t,t',X]
\cos(2Rt')
\end{equation}
and
\begin{equation}
F_\alpha[t,t',X]=\cos( 2X\cos t \cos t') \cos(2\alpha t) .
\end{equation}
For both of these terms the integral over $t$
can be  performed using
 3.715.19 from ref. \cite{Grad},
 and the integral over $t'$ can then be
performed using 6.681.1 from ref.
\cite{Grad}.  The end result is the desired quantity
\begin{equation}
W_\alpha = (-1)^\alpha \left[ -\frac{J_\alpha^2(X)}{2R^2} +
\frac{\pi J_{\alpha+R}(X)J_{\alpha-R}(X)}{2R\sin(R\pi)} \right]
\end{equation}

Using this partial result we can calculate the more
 relevant quantity
\begin{eqnarray}
\tilde W_\alpha
&=& \sum_{n=1}^{\infty} \frac{n^2 }{R^2 - n^2}  J_{n+\alpha}(X)
 J_{n-\alpha}(X) \\
&=& \sum_{n=1}^{\infty} \left[-1 + \frac{R^2}{R^2 - n^2}
 \right] J_{n+\alpha}(X)
 J_{n-\alpha}(X)
\end{eqnarray}
The first term can be evaluated using the orthogonality relation of
Bessel functions (Eq. 9.1.75 ref. \cite{Abrom})
and the second term is just $R^2 W_\alpha$.
Thus we have
\begin{equation}
\tilde W_\alpha = -\frac{1}{2}\delta_{\alpha,0} + (-1)^\alpha
\frac{\pi R }{2\sin(R\pi)} J_{\alpha+R}(X)J_{\alpha-R}(X)
\end{equation}

Now using the $\alpha = 0$ case of this result immediately
allows us to
perform one of the desired sums giving us the result
\begin{equation}
    \label{eq:xxend}
    \tilde \sigma_{xx} = \frac{ipe^2}{\pi \hbar} \frac{2R}{X^2}
 \left[
-\frac{1}{2} + \frac{\pi R}{2 \sin(R \pi)} J_R(X) J_{-R}(X) \right]
\end{equation}
Furthermore, by differentiating $\tilde W_\alpha$ with respect
 to $X$, we can derive
\begin{equation}
 \tilde \sigma_{xy} = \frac{pe^2}{\pi \hbar}
    \frac{\pi R}{2 X\sin(R\pi)} \left[J_R(X) J_{-R}(X)\right]'
\end{equation}
By using the Bessel function identities 9.1.27
 from ref. \cite{Abrom}
as well as the Wronskian identity 9.1.15 from ref. \cite{Abrom}
we can rewrite the off diagonal conductance as
\begin{equation}
\label{eq:xyend}
 \tilde \sigma_{xy} = -i \tilde \sigma_{xx}-\frac{pe^2}{\pi \hbar}
\frac{R\pi}{X\sin(R\pi)} J_{R+1}(X) J_{-R}(X)
\end{equation}

The evaluation of $\tilde \sigma_{yy}$ is achieved
by using the same
 Bessel function
identities 9.1.27 from ref. \cite{Abrom} to derive
\begin{equation}
 [J_n'(X)]^2 = -J_{n-1}(X)J_{n+1}(X) +\frac{n^2}{X^2} J_n^2(X) .
\end{equation}
The sum over the first term
is just $W_1$ whereas the sum over the second
term is $X^{-2} \tilde W_0$.  The result is easily simplified to
\begin{equation}
\label{eq:yyend}
\tilde \sigma_{yy} = \tilde \sigma_{xx} +  \frac{ipe^2}{\pi
\hbar}\frac{\pi}{\sin(R\pi)} J_{1+R}(X)J_{1-R}(X).
\end{equation}

We can evaluate some of the limits of this expression for the
conductivity by expanding the Bessel functions in their defining
series (Eq. 9.1.10 of ref.  \cite{Abrom}).  The condition for this
series expansion to be a good approximation is that $X^2/R$ be
much less than one.  If we insert this expansion into the above
expressions (and using equation 6.1.17 of ref. \cite{Abrom}) to
simplify the result, we find the expected result
\begin{equation}
   \tilde \sigma = \frac{e^2 n_e}{m_b}
 \frac{1}{(\Delta \omega_c){^2}
     - \omega^2}
    \left[  \begin{array}{cc} -i\omega & -\Delta \omega_c \\
                               \Delta \omega_c & -i\omega
\end{array}
            \right] + {\cal O}(X^2/R)
\end{equation}
The important thing to realize here is that within the
semiclassical approximation, a low $q$ expansion and a large
$\omega$ expansion are equivalent.  This is not obvious from the
original expression for the conductivity
(Eq.(\ref{eq:sigmatilde})), but becomes clear once we have this
closed form expression.
\section{Fermi-Liquid Theory}
\label{app:FLT}

In this appendix we use the Landau-Silin Fermi-liquid
theory\cite{FermiLiquid,Lee,Pines} to determine the effect of
mass renormalization on the conductivity of a system in a
magnetic field.  We assume here a  two-dimensional system  of
spinless Fermions in a magnetic field $B$.  (Note that in the
text we consider a system in a field $\Delta B$.  We have dropped
the delta for simplicity of notation.) For this system, the
linearized semiclassical transport equation (which should be
accurate in the semiclassical regime as discussed in Sec.
\ref{sec:mean}) is written as\cite{FermiLiquid,Pines}
\begin{eqnarray} \nonumber
    \frac{\partial \delta n}{\partial t}
&+& \left[ \vec v_{\vec k} \cdot
\vec \nabla_{\vec r} -
\frac{e}{c} (\vec v_{\vec k} \times \vec B)\cdot
 \vec \nabla_{\vec k}
 \right] [ \delta n
+ \delta \epsilon_1 \frac{\partial n_0}{\partial \epsilon_0}] =
\\   \label{eq:trans1}
&-& e \vec E \cdot \vec v_{\vec k} \frac{\partial n_0}{\partial
\epsilon_0} + I
\end{eqnarray}
where $\vec k = \vec p + \frac{e}{c} \vec A$ is the kinetic
momentum, $\epsilon_0 = k^2/(2m^*$) is the kinetic energy of a
noninteracting quasiparticle, $m^*$ is the effective mass of a
quasiparticle at the Fermi surface, $\delta n(\vec k,\vec r)$ is
the local deviation from the equilibrium distribution $n_0(\vec
k)$, the local quasiparticle velocity is given by $\vec v_{\vec
k} = \nabla_{\vec k } (\epsilon_0 + \delta \epsilon_1)$, the
effects of scattering are included in the collision integral
$I(n_0 + \delta n)$, and the local change in quasiparticle energy
$\delta \epsilon_1$ due to interaction  is given by
\begin{equation}
    \label{eq:dep}
    \delta \epsilon_1(\vec k,\vec r) =
\frac{1}{(2\pi)^2} \int d^2  k \,'
\Phi(\vec k, \vec k\,') \delta n(\vec k\,',\vec r)
\end{equation}
where $\Phi(\vec k, \vec k\,')$ is the (unknown) Landau
interaction function.  The above transport equation
(\ref{eq:trans1}) is derived (as described loosely in
\ref{sub:ModSem}) by using a single particle effective
Hamiltonian derived from a local energy functional and then using
Hamilton's equations of motion for this effective
Hamiltonian\cite{FermiLiquid,Pines}.  The equations of motion
must be expanded to linear order in the effects of the perturbing
electromagnetic field to yield Eq. (\ref{eq:trans1}).

Although the inclusion of a nonzero scattering integral is
straightforward\cite{Lee}, we will assume that $I \rightarrow 0$
($\tau \rightarrow \infty$)   for simplicity. Keeping with our
convention that $\vec q \| \hatn x$, we apply an electromagnetic
perturbation proportional to $e^{iqx-i\omega t }$.  Following Lee
and Quinn\cite{Lee} we rewrite the linearized transport equation
(\ref{eq:trans1}) as\cite{FermiLiquid,Lee}
\begin{equation}
\label{eq:kin1}
-i\omega f(\phi) +
\left(iqv_x(\phi) +
 \omega_c^* \frac{\partial}{\partial \phi} \right)
[f(\phi) + \delta \epsilon_1(\phi)] = -e \vec E \cdot
\vec v(\phi) .
\end{equation}
where $\omega_c^* = eB/(m^* c)$ is the mass renormalized  cyclotron
frequency, $f(\phi)$ is defined by
\begin{equation}
\delta n(\vec k) =  f(\phi) \frac{-\partial n_0}{\partial
\epsilon_0(\vec k)}
\end{equation}
with $\phi$ is the angle defining the direction of
 the kinetic momentum
 $\vec k$,  and the
velocity vector is given by
\begin{equation}
    \vec v(\phi) = v_{\mbox{\tiny{F}}} (\sin \phi, -\cos \phi).
\end{equation}
with $ v_{\mbox{\tiny{F}}} =  k_{\mbox{\tiny{F}}}/m^*$.
 Since $f$
is periodic in $\phi$ we can expand it in a Fourier series
\begin{eqnarray}
    f_l  &=& \frac{1}{2\pi} \int_0^{2\pi}
 d\phi f(\phi) e^{il\phi} \\
    f(\phi) &=& \sum_{l} e^{-il\phi} f_l.
\end{eqnarray}
 Similarly, $\Phi(\vec k, \vec k\,')$
is periodic in $\phi - \phi'$ so we can write
\begin{equation}
        \label{eq:Phi1}
    \Phi(\vec k, \vec k\,') = \frac{2\pi}{m^*} \sum_{l} A_l
e^{-il(\phi - \phi')}.
\end{equation}
 Furthermore, since $\Phi$
is symmetric and real, we have $A_l = A_{-l}$ and $A_l$ is real.

We can define a displacement vector for
the quasiparticles on the Fermi surface
\begin{equation}
     \vec R(\phi) =- \frac{1}{\omega_c^*}
 \int_{\pi/2}^{\phi} \vec v(\phi') d\phi' .
\end{equation}
so that\cite{Grad}
\begin{equation}
    e^{i\vec q \cdot \vec R(\phi)} =
e^{iX\cos \phi} = \sum_n i^n J_n(X) e^{in\phi}
\end{equation}
for $X = qR_c^* = q
 v_{\mbox{\tiny{F}}}/\omega_c^*$ and $J_n$ is the Bessel
function.
We now expand the following periodic functions of
 $\phi$ into Fourier series
\begin{eqnarray}
        f(\phi)  e^{-i\vec q \cdot \vec R(\phi)}
 &=& \sum_{l} e^{-il\phi} F_l
\\
\delta \epsilon_1(\phi) e^{-i\vec q \cdot \vec R(\phi)}
 &=& \sum_{l} e^{-il\phi} \epsilon_l
\\
\vec v(\phi) e^{-i\vec q \cdot \vec R(\phi)}  &=&
\sum_{l} e^{-il\phi} \vec v_l
\end{eqnarray}
By inverting these Fourier series and
inserting into above definitions we
derive the following relations
\begin{eqnarray}
\label{eq:efromf}
    \epsilon_l &=& \sum_m A_m f_m i^{m-l} J_{l-m}(X) \\
\label{eq:ffromF}
    f_l &=& \sum_m i^{m-l} J_{m-l}(X) F_m \\
\label{eq:vfromJ}
    \vec v_l &=& - v_{\mbox{\tiny{F}}} i^{-l}
        \left( \begin{array}{c} lJ_l(X)/X \\ -iJ'_l(X)\end{array}
\right)_.
\end{eqnarray}
where the Bessel function identities 9.1.27 from ref. \cite{Abrom}
 have been used to
derive $\vec v_l$.   Note that the coefficients $\vec v_l$ are
the velocity coefficients used in Eq.
(\ref{eq:sigmatilde}) to calculate the quasiparticle
 conductivity  up to
multiplicative constants.  By using
the Bessel function orthogonality relation 9.1.75 from ref. \cite{Abrom}
 we can
also derive
the inverse relation
\begin{equation}
    \label{eq:Ffromf}
    F_m  = \sum_l i^{l-m}J_{m-l}(X) f_l.
\end{equation}

Now multiplying our kinetic equation (\ref{eq:kin1}) by
$(2\pi)^{-1}\exp({il\phi-i\vec q\cdot \vec R(\phi)})$ and
integrating over
$\phi$ yields the kinetic equation in terms of our new variables
(previously derived by Lee and Quinn\cite{Lee})
\begin{equation}
\label{eq:newkin}
    i\omega F_l + il\omega_c^*[F_l + \epsilon_l] =
e\vec E\cdot \vec v_l
\end{equation}
where we have used the fact that
\begin{equation}
i q v_x e^{-i\vec q\cdot \vec R(\phi)} =
\omega_c^* \frac{d}{d\phi} e^{-i\vec q\cdot \vec R(\phi)}
\end{equation}
in our evaluation of the
integral.

We now want to
express the current in terms of the motion of quasiparticles.
This can be
done using the standard result of Fermi-liquid
theory\cite{FermiLiquid,Pines}
\begin{eqnarray} \nonumber
    \vec j &=& \frac{-e}{(2\pi)^2}
\int d^2k \, \delta n(\vec k) \mbox{{\huge [}} \vec v(\vec k) + \\
& & \frac{1}{(2\pi)^2} \int d^2 k' \Phi(\vec k,\vec k\,')
\vec v(\vec k\,')
\left(\frac{-\partial n_0}{\partial
\epsilon(\vec k\,')} \right) \mbox{{\huge ]}}
\end{eqnarray}
where the second term represents the backflow current due
to interactions.
By interchanging the order of integration we can rewrite this as
\begin{equation}
 \vec j = \frac{-e}{(2\pi)^2} \int d^2k \vec v(\vec k)
\left(\frac{-\partial n_0}{\partial \epsilon(\vec k)} \right)
(f(\vec k) + \delta \epsilon_1(\vec k)).
\end{equation}
which can be expressed in terms of our new variables as
\begin{equation}
     \vec j = \frac{-em^*}{2\pi} \sum_l \vec v_l^{\,*}
 (F_l + \epsilon_l).
\end{equation}
Combining this with Eq. (\ref{eq:newkin}) yields the result
\begin{equation}
    \label{eq:jfF}
     \vec j = \frac{-em^*}{2\pi} \sum_l \vec v_l^{\,*}
\left[\frac{ e\vec E\cdot \vec v_l - i\omega F_l}{i l \omega_c^* }
 \right]_.
\end{equation}

At this point let us consider what happens in a noninteracting
system.  In this case, all  the Fermi liquid coefficients $A_l$
and hence $\epsilon_l$ are zero.  The kinetic equation
(\ref{eq:newkin}) is solved by
\begin{equation}
    \label{eq:Fnon}
    F_l = \frac{ e\vec E\cdot \vec v_l}{i\omega + il\omega_c^*}
\end{equation}
so that the noninteracting  current is given by
\begin{equation}
    \label{eq:jfdef}
    \vec j^{\mbox{\small n}} =
 \frac{-e^2 m^*}{2\pi} \sum_l \left[ \frac{1}{i\omega +
il\omega_c^*}\right] \vec v_l^{\,*} \:\: \vec v_l \cdot \vec E
\end{equation}
and thus the conductivity is
\begin{equation}
    \label{eq:sfdef}
    \sigma^{\mbox{\small n}}_{ij} =  \frac{-e^2 m^*}{2\pi}
 \sum_l \left[ \frac{1}{i\omega +
il\omega_c^*}\right] (\vec v_l^{\,*})_j (\vec v_l)_i
\end{equation}
which is exactly the semiclassical expression
\cite{Tome,Harrison} for the
conductivity of a system of noninteracting quasiparticles
in the $\tau \rightarrow \infty$ limit
 given in  Eq. (\ref{eq:sigmatilde}).

We now want to analyze this system when the interaction
coefficients are nonzero.  To do this, we must be able to solve
the kinetic equation (\ref{eq:newkin}).  We express the kinetic
equation in terms of the unknown variables $f_l$ by using Eqs.
(\ref{eq:efromf}) and (\ref{eq:ffromF}) to yield the matrix
equation\cite{Lee}
\begin{equation}
\label{eq:sys}
 c_n = \sum_m (a_n^m - \delta_{nm})  f_m
\end{equation}
where
\begin{eqnarray}
    \label{eq:cdef}
    c_n &=& -\sum_l i^{n-l} \frac{e \vec E \cdot \vec v_l J_{l-n}}
{i\omega + il\omega_c^*}
  \\  \label{eq:adef}
    a_n^m &=& i^{n-m} A_m \left[-\delta_{nm} + \sum_l
 \frac{i\omega J_{l-m} J_{l-n}}{i\omega + il\omega_c^*} \right]
\end{eqnarray}
where the Bessel functions and the velocity coefficients $\vec
v_l$ are evaluated at $X$.  Although this system of equations is
infinite dimensional, if we assume that $A_i$ is zero for $i$
greater than some number $i_{max}$, then we have $a^i_j$ also
zero for $i>i_{max}$.  In this case the equations with $ -i_{max}
\le n \le i_{max}$ form a closed system of equations with
variables $\{ f_{(-i_{max})} \ldots f_{(i_{max})} \}$ where $f_0$
is real and all other $f_n$ are complex.  Once this smaller
system is solved, the remaining $f_n$ are defined trivially since
they only depend on the already determined values.  Then one can
solve for the $F_n$ using Eq. (\ref{eq:Ffromf}) and then find the
current using Eq. (\ref{eq:jfF}) and hence extract the
conductivity.

As an illustrative example we consider the case where $A_1$ is
the only nonzero Fermi liquid coefficient and using this
approximation (whose validity is discussed in Sec.
\ref{sub:ModSem}) we derive the same result (Eq.
(\ref{eq:modsem})) as in  Sec. \ref{sub:ModSem}.    Note that in
Sec.  \ref{sub:ModSem} we use a trick to perform this same
calculation that can not be generalized to account for an
arbitrary number of nonzero Fermi-liquid coefficients.   The
method shown below is more difficult, but more generalizable (in
principle one could also generalize this method to include the
effects of impurity scattering also).

With the simplification that $A_1$ is the only nonzero
Fermi-liquid coefficient, we now have the decoupled system of two
 equations
\begin{eqnarray} \label{eq:pair}  \begin{array}{cc}
    c_{1} =& (a_{1}^{1} -1) f_{1} + a_{1}^{-1} f_{-1} \\
    c_{-1} =& (a_{-1}^{-1} -1)f_{-1} + a_{-1}^{1} f_{1}  .
\end{array}
\end{eqnarray}
Solving this system yields the result
\begin{equation}\begin{array}{cc}  \label{eq:fans}
    f_{1} =& D^{-1} [ (a^{-1}_{-1} -1) c_1 - a_{1}^{-1} c_{-1}] \\
    f_{-1} =& D^{-1} [(a^{1}_1 -1) c_{-1} - a_{-1}^1 c_1 ]  \\
    D =& (a_{-1}^{-1} -1)(a_{1}^{1} -1) -a_{1}^{-1} a_{-1}^1 .
\end{array}
\end{equation}
Once $f_1$ and $f_{-1}$ are determined, all of the $f_n$
are then given by the Eq. (\ref{eq:sys}) which now takes the form
\begin{equation}
    f_n = a_n^{1} f_1 + a_n^{-1} f_{-1} - c_n .
\end{equation}
Using this result in  Eq. (\ref{eq:Ffromf}), inserting
the definition
of $c_n$, and simplifying by using the
Bessel function orthogonality equation (Eq. 9.1.75 ref.
\cite{Abrom})  yields
\begin{equation}
    F_l = \frac{ e\vec E\cdot \vec v_l}{i\omega + il\omega_c^*} +
 \frac{A_1 i^{l} l\omega_c^*}{i\omega+ il\omega_c^*}
[f_{-1} J_{m+1} - f_1 J_{m-1}] .
\end{equation}
Notice that the first term is just the noninteracting result
given in Eq.  (\ref{eq:Fnon}), whereas the second term is clearly
an interaction term.  Substituting this expression into Eq.
(\ref{eq:newkin})and using Bessel function identities ( 9.1.27
from ref. \cite{Abrom}) yields the current
\begin{equation}
    \vec j = \vec j^{\mbox{\small n}} +\delta \vec j
\end{equation}
where  $\vec j^{\mbox{\small n}}$ is the previous noninteracting
current defined in Eq. (\ref{eq:jfdef}) and
\begin{eqnarray}
\delta \vec j &=&
\frac{-\omega e m^* A_1}{2 \pi} \sum_l \frac{i^l \vec
v^{\,*}_l}{i\omega + i\omega_c^*} \\ \nonumber
&\times&  \left[ f_1 \left(\frac{lJ_l}{X}
 +J_l' \right) -
f_{-1} \left( \frac{lJ_l}{X} - J_l' \right) \right]_.
\end{eqnarray}
By using the definition of  $\vec v_l$ in terms of Bessel
functions as given in Eq.
(\ref{eq:vfromJ}), we can put this in the simple form
\begin{equation}
    \delta \vec j = \frac{\omega A_1}{v_{\mbox{\tiny{F}}} e}
 [f_{-1}
\sigma^{\mbox{\small n}} \hatn r_- - f_1
\sigma^{\mbox{\small n}} \hatn r_+]
\end{equation}
where $\hatn r_+ = \hatn r_-^* = \hatn x +
i\hatn y$ and $\sigma^{\mbox{\small n}}$ is
the previous noninteracting
conductivity defined in Eq.
(\ref{eq:sfdef}).
 It should be noted that the coefficients $f_i$ are linear in the
$c_i$'s which in turn are linear in $\vec E$ as can be seen from
Eqs (\ref{eq:fans}) and (\ref{eq:cdef}).  Hence $\delta \vec j$
and $\vec j$ will be linear in the field $\vec E$ such that a
linear conductivity can be defined properly.

At this point a great deal of very tedious algebra (along with
clever use of the definition of $\sigma^{\mbox{\small n}}$ in
terms of Bessel functions) can be used to simplify the result
into the form given in Eq.  (\ref{eq:modsem}).

\newpage
\bigskip
\bigskip

{{\large{\bf \sc Figure Captions}}}

\bigskip

\begin{figure} \caption{} {The location and weights of the poles
in the response function $ K_{00}(q,\omega)$ calculated in the
semiclassical approximation (Eqs. (\ref{eq:sigmatilde}), and
(\ref{eq:SemRPA1}) - (\ref{eq:rtfromst})) and using the bare band
mass $m_b$, a vanishing Coulomb potential $\epsilon \rightarrow
\infty$, and no scattering ($\tau \rightarrow \infty$).  Results
are shown for filling fractions $\nu = \frac{1}{3},\frac{3}{7}$,
and $\frac{10}{21}$ corresponding to effective Landau level
fillings $p=1,3$, and $10$ respectively where $m=1$.  Solid curves
show the locations of the poles; the width of the striped band
around each pole is $q^{-2}$ times the weight of the pole in
$K_{00}$.  Note that the cyclotron mode ($\omega = (2p+1) \Delta
\omega_c = \omega_c$) has all of the weight at $q \rightarrow 0$
in accordance with Kohn's theorem.} \label{fig:sm1v0} \end{figure}

\begin{figure} \caption{}{The semiclassical orbiting mode of a
quasiparticle in a perturbing field $\vec E_{\mbox{\small eff}}$
that represents the sum of the Chern-Simons and actual
electromagnetic  fields.  Here we show the effect of applying a
zero frequency wave with wavelength less than the effective
cyclotron diameter to an orbiting quasiparticle.  As described in
the text, when the quasiparticle is moving essentially parallel to
the wavevector (shown dotted) it experiences an oscillating field.
However, when the quasiparticle is moving perpendicular to the
wavevector (shown solid) then it feels the same force for a long
period of time.} \label{fig:orbit1} \end{figure}

\newpage

\begin{figure}\caption{} {Semiclassical orbiting modes of a
quasiparticle in an electromagnetic wave at nonzero frequency.
Here we show the world lines of oscillating particles (sinusoidal)
and the world lines (parallel lines) of the wave crests.  As
described in the text, in order to have a zero in the
quasiparticle conductivity and hence a pole in the electromagnetic
response, we must arrange so that the phase of the wave is the
same on the extreme left as it is on the extreme right so that the
largest contributions to the integral  (\ref{eq:sigint}) cancel.
In Case I, each time the quasiparticle moves to the right  it
begins and ends at the same phase of the wave.  In Case II, each
time the quasiparticle moves left it begins and ends at the same
phase of the wave.  These two cases are not, in general,
equivalent since the phase of the wave is not the same at the
beginning and the end of an orbit.  \label{fig:world} }
\end{figure}

\begin{figure} \caption{}{The location and weights of the poles in
the response function $K_{00}(q,\omega)$ calculated in the RPA
(Eqs.  (\ref{eq:K0def}), (\ref{eq:sfree}), (\ref{eq:Sigmadef}))
 using the bare band
mass $m_b$ and a vanishing Coulomb potential  ($\epsilon
\rightarrow \infty $).  Results are shown for filling fractions
$\nu = \frac{1}{3},\frac{3}{7}$, and $\frac{10}{21}$ corresponding
to effective Landau level fillings $p=1,3$, and $10$ respectively
where $m=1$.  The width of the striped band around each pole
(solid)  is $q^{-2}$ times the weight of the pole in $K_{00}$.  In
the large $p$ limit our results look very much like the
semiclassical results of Figure \ref{fig:sm1v0}.  }
\label{fig:rm1v0} \end{figure}

\begin{figure} \caption{}{The location and weights of the poles in
the response function $ K_{00}(q,\omega)$ calculated in the
modified RPA  as described in Sec.  \ref{sub:modRPA} using a
renormalized mass $m^* = 3.9 m_b$ and a Coulomb potential for an
electron density $n_e = 10^{-11} \mbox{cm}^{-2}$ and a dielectric
constant $\epsilon = 12.6$. Results are shown for filling
fractions $\nu = \frac{1}{3},\frac{3}{7}$, and $\frac{10}{21}$
corresponding to effective Landau level fillings $p=1,3$, and $10$
respectively  where $m=1$.  The width of the striped band around
each pole (solid)  is $q^{-2}$ times the weight of the pole in
$K_{00}$.  This figure is the complete theory including both
Coulomb interaction and mass renormalization from Fermi liquid
interaction term $A_1$.  The quantity $\Delta \omega_c^*$ is
defined as $e|\Delta B|/(m^*c)$, where $\Delta B = B/(2p+1)$.  }
\label{fig:rm3vfinal} \end{figure}

\end{document}